\documentclass[twocolumn,showpacs,preprintnumbers,amsmath,amssymb,aps,prb,longbibliography]{revtex4-2}
\usepackage{graphicx}% Include figure files
\usepackage{dcolumn}% Align table columns on decimal point
\usepackage{bm}% bold math
\usepackage{verbatim} %needed for multiline comments
\usepackage[usenames, dvipsnames]{color} %needed for comment colors
\usepackage{ulem}
\usepackage{color}
\usepackage[colorlinks=false,pdfencoding=auto,psdextra]{hyperref}
\usepackage{multirow}
\usepackage{tabularx}
\usepackage{dsfont}

\definecolor{orange}{rgb}{1.0, 0.5, 0.0}

\begin{document}
\title{Mode locking of hole spin coherences in \\ {CsPb(Cl,Br)}$_3$ perovskite nanocrystals %in glass matrix
}

%Authors of paper
\author{E. Kirstein$^{1}$, N.~E. Kopteva$^{1}$, D.~R. Yakovlev$^{1,2,3}$, E.~A. Zhukov$^{1,2}$, E.~V. Kolobkova$^{4,5}$, \\ M.~S. Kuznetsova$^{6}$, V.~V. Belykh$^{3}$, I.~A. Yugova$^{6}$, M.~M. Glazov$^{2}$, M.~Bayer$^{1,2}$, and A. Greilich$^{1}$}

\affiliation{$^{1}$Experimentelle Physik 2, Technische Universit\"at Dortmund, 44227 Dortmund, Germany}
\affiliation{$^{2}$Ioffe Institute, Russian Academy of Sciences, 194021 St. Petersburg, Russia}
\affiliation{$^{3}$P. N. Lebedev Physical Institute of the Russian Academy of Sciences, 119991 Moscow, Russia}
\affiliation{$^{4}$ITMO University, 199034 St. Petersburg, Russia}
\affiliation{$^{5}$St. Petersburg State Institute of Technology, 190013 St. Petersburg, Russia}
\affiliation{$^{6}$Spin Optics Laboratory, St. Petersburg State University, 198504 St. Petersburg, Russia}

%\date{\today}

\begin{abstract}
The spin physics of perovskite nanocrystals with confined electrons or holes is attracting increasing attention, both for fundamental studies and spintronic applications. Here, stable CsPb(Cl$_{0.5}$Br$_{0.5}$)$_3$ lead halide perovskite nanocrystals embedded in a fluorophosphate glass matrix are studied by time-resolved optical spectroscopy to unravel the coherent spin dynamics of holes and their interaction with nuclear spins of the $^{207}$Pb isotope. We demonstrate the spin mode locking effect provided by the synchronization of the Larmor precession of single hole spins in each nanocrystal in the ensemble that are excited periodically by a laser in an external magnetic field. The mode locking is enhanced by nuclei-induced frequency focusing. An ensemble spin dephasing time $T_2^*$ of a nanosecond and a single hole spin coherence time of $T_2=13\,$ns are measured. The developed theoretical model accounting for the mode locking and nuclear focusing for randomly oriented nanocrystals with perovskite band structure describes the experimental data very well.
\end{abstract}

\maketitle

%\cD{NN rules: Articles: an abstract of approximately 150 words, unreferenced; main text of no more than 3,000 words and 6 display items (figures, tables). As a guideline, Articles allow up to 50 references. Section headings should be used and subheadings may appear in 'Results'. Avoid 'Introduction' as a heading.}

Lead halide perovskite semiconductors are highly attractive due to their remarkable photovoltaic efficiency~\cite{jena2019,nrel2021}, and are promising for optoelectronic~\cite{Vinattieri2021_book,Vardeny2022_book} and spintronic~\cite{Vardeny2022_book,wang2019,ning2020,kim2021} applications. Perovskite nanocrystals (NCs) have recently extended the wide class of semiconductor NCs grown by colloidal synthesis~\cite{Kovalenko2017,Chen2018a,Protesescu2015,Akkerman2018,Park2015,Yu2021}. They show a remarkable quantum yield of up to 90\% even for bare NCs, as surface states do not act detrimentally on the exciton emission efficiency. For inorganic CsPb$X_3$ ($X = $ I, Br, or Cl) NCs, the band gap can be tuned from the infrared up to the ultraviolet by mixing the halogen composition and by changing NC size, varying the quantum confinement of charge carriers. Additionally, the exciton fine structure can be adjusted by the NC shape~\cite{Becker-Nature2018,Lin2016,Tamarat2019,Ramade2018}.

In addition to the high quantum yield and tunable optical properties, the simple fabrication makes lead halide perovskite NCs interesting for applications. As far as spintronics is concerned, only quite a few studies have been performed so far. Neutral and charged excitons in single NCs were identified by their Zeeman splitting in magnetic field~\cite{Fu2017,Isarov2017}. For NC ensembles, negatively charged excitons (trions) and dark excitons were identified in strong magnetic fields of 30\,T~\cite{Canneson-NanoLett2017}. Optical orientation, optical alignment and anisotropic exciton Zeeman splitting were observed~\cite{Nestoklon-PRB2018}. The coherent spin dynamics of electrons and holes in CsPbBr$_3$ NCs~\cite{Crane2020,Grigoryev2021} were explored, and the picosecond spin dynamics of carriers in CsPbI$_3$ NCs were reported~\cite{Strohmair2020}. Despite the progress in crystal growth, perovskite NCs still suffer from insufficient long-term stability of optical properties. A promising approach here is to synthesize NCs embedded in glass, providing protection by encapsulation~\cite{chen2018b,kolobkova2021}.

As a result, perovskite NCs in a glass matrix may be suitable for quantum technologies. For reference, one may compare NCs with the established system of self-assembled (In,Ga)As/GaAs quantum dots (QDs) singly charged with an electron or a hole. Using all-optical approaches one can orient and manipulate the spins of the charge carriers on the picosecond timescale at high operation frequencies~\cite{yamamoto2008,greilich2009}. There are two complementary concepts: the first one uses a single spin in a QD for establishing a quantum bit, the second one exploits a QD ensemble which may be suited for a quantum memory with sufficiently strong light-matter interaction. Here we focus on a NC ensemble, which gives information on the average spin properties and their scatter due to inhomogeneity. The inhomogeneity obstacle, however, can be overcome by applying periodic laser excitation that synchronizes the electron spin Larmor precession  in different NCs subject to a transverse magnetic field. This spin mode locking (SML) effect, initially discovered in singly charged (In,Ga)As QDs, allows one to uncover the spin coherence of individual carriers, which is typically prevented due to the faster spin dephasing in the ensemble~\cite{greilich2006,greilich2007,yakovlevCh6}. The interaction of resident carrier spins with the surrounding nuclear bath provides further flexibility with potential memory times up to hours and exchange interaction fields up to a few Tesla~\cite{evers2021a,evers2021b}. This collective phenomenon of spin dynamics homogenization has been demonstrated so far only for (In,Ga)As QDs.

In this paper we demonstrate the SML effect for holes in a totally different class of QDs, namely perovskite CsPb(Cl$_{0.5}$Br$_{0.5}$)$_3$ NCs synthesized in a glass matrix. We provide detailed information on the spin dynamics of confined holes correlated with the appearance of spin mode locking. We measure the $g$-factor, spread of $g$-factors, longitudinal spin relaxation time $T_1$, transversal spin coherence time $T_2$, and inhomogeneous spin dephasing time $T_2^*$. Exploiting the hyperfine interaction with the nuclear spins, we implement dynamic nuclear polarization and optically-detected nuclear magnetic resonance (ODNMR) to identify the involved nuclear isotopes. A theoretical model of the SML in the perovskite is developed to account for the inverted band structure and for the dominating role of the hole-nuclear interaction, compared to self-assembled QDs. As an additional challenge, the random orientation of NCs in the ensemble is considered.

\section*{Results}

\begin{figure*}[t!]
\begin{center}
\includegraphics[width=17cm]{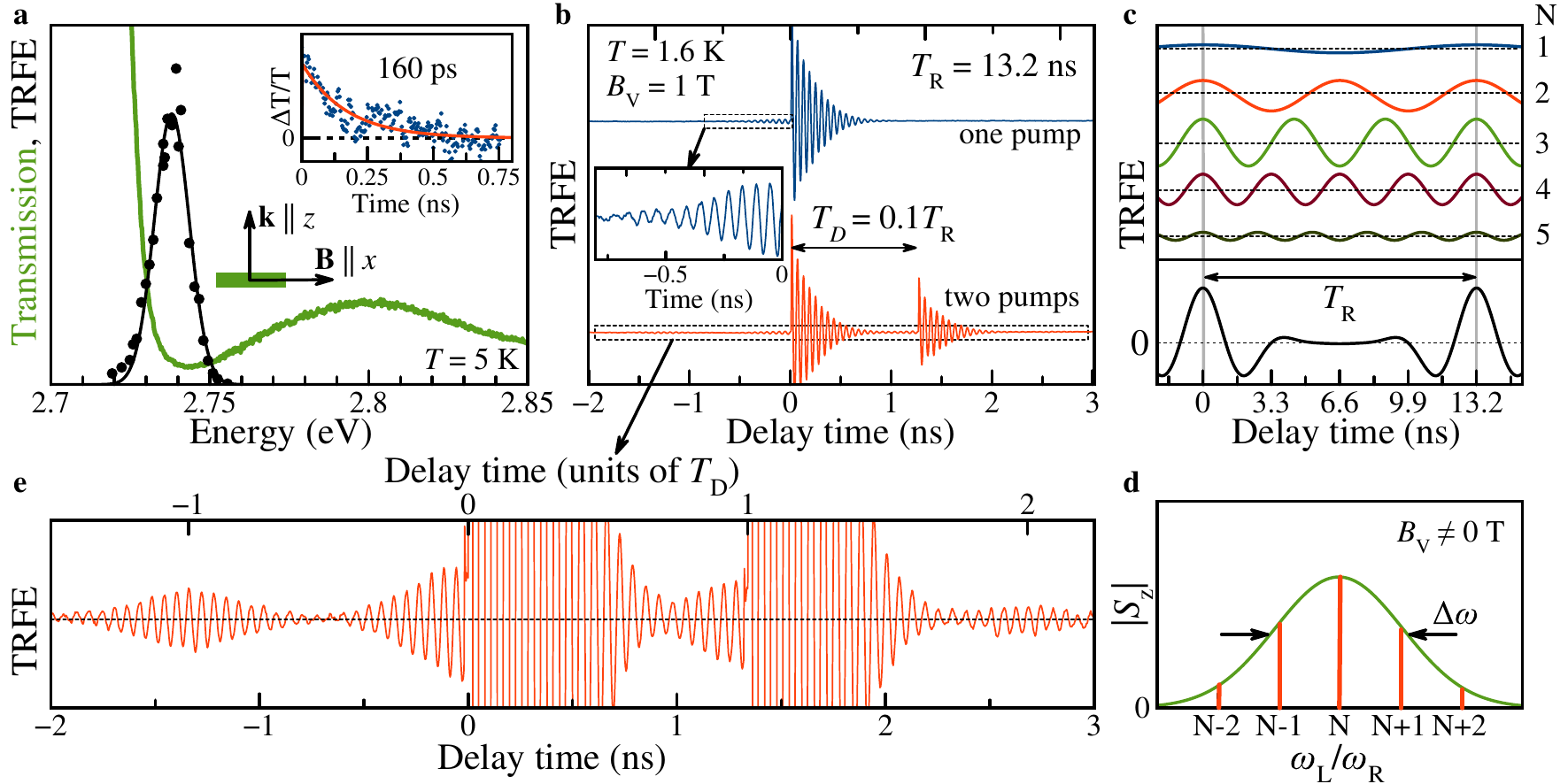}
%includegraphics[trim=0mm 0mm 0mm 0mm, clip, width=2.05\columnwidth]{Fig02.pdf}
\caption{\label{fig1} \textbf{Spin mode locking in CsPb(Cl,Br)$_{3}$ NCs.}
\textbf{a}, Transmission spectrum of CsPb(Cl$_{0.5}$Br$_{0.5}$)$_3$ NCs (green line). Black dots show spectral profile of the Faraday ellipticity amplitude measured at zero pump-probe time delay fitted by a Gaussian function (black line). $T = 5$\,K. Inset shows time-resolved differential transmission (dots) with a monoexponential fit (red line) providing the exciton lifetime of 160\,ps.
\textbf{b}, Time-resolved Faraday ellipticity (TRFE) measured at the photon energy of 2.737\,eV with 1.5\,ps laser pulses. Blue curve shows data for the one-pump protocol with a repetition period of $T_\text{R} = 13.2$\,ns using the pump power $P_\text{pu} = 25$\,mW. Inset shows zoom of signal before the arrival of the pump pulse, indicated by the box. Red curve is TRFE for the two-pump protocol ($T_\text{D} = 0.1T_\text{R} = 1.32$\,ns) using the pump powers $P_\text{pu,1} = 25$\,mW and $P_\text{pu,2} = 13$\,mW. $T = 1.6$\,K and $B_{\rm V} = 1$\,T.
\textbf{c}, Scheme shows spins precessing at five lowest PSC mode frequencies. Black curve at the bottom shows sum signal of these modes, each weightes assuming an ensemble with a Gaussian distribution of amplitudes, as shown in panel d by the red vertical lines.
\textbf{d}, Illustration of the pumped carrier spin polarizations in inhomogeneous NC ensemble subject to a magnetic field. Green line shows the distribution of carrier precession frequencies caused by dispersion of the Larmor frequency. It is modeled by Gaussian with width $\Delta \omega$. PSC modes for the one-pump protocol, fulfilling the condition $\omega_{\rm L}=N \omega_\text{R} = 2\pi N/T_\text{R}$, are shown by the red vertical lines.
\textbf{e}, Zoom of the TRFE signal in the two-pumps protocol from panel b, indicated by the box. Upper scale gives delay in units of separation, $T_{\rm D}$, between two pumps. The bursts in the signal are associated with the electron spins with Larmor frequencies that are commensurate with the frequency: $\omega_\text{D} = 2\pi M/T_\text{D}$.
}
\end{center}
\end{figure*}

The optical properties of CsPb(Cl$_{0.5}$Br$_{0.5}$)$_3$ nanocrystals embedded in a fluorophosphate glass are shown in Fig.~\ref{fig1}a. At cryogenic temperature of $T=5$~K the transmission spectrum of the NCs with about 7~nm size shows pronounced exciton resonance at 2.743~eV, which is broadened due to NC size dispersion.  The exciton resonance can be also traced by the spectral dependence of the time-resolved Faraday ellipticity (TRFE) amplitude. The exciton population dynamics is measured by time-resolved differential transmission ($\Delta T/ T$), showing a decay time of 160~ps, which corresponds to the exciton lifetime (inset in Fig.~\ref{fig1}a).

To study the coherent spin dynamics of resident carriers we use the pump-probe technique with spin-sensitive detection of the Faraday ellipticity, whose potential has been approved for lead halide perovskite crystals~\cite{belykh2019,kirstein2021,kirstein2021nc}, polycrystalline films~\cite{odenthal2017,garcia-arellano2021,Grigoryev2021}, and CsPbBr$_3$ NCs~\cite{Crane2020,Grigoryev2021}. The pump laser pulses are circularly polarized and, according to the selection rules, generate spin polarization of electrons and holes along the light $\mathbf{k}$ vector. The oriented carrier spins precess at the Larmor frequency around the magnetic field applied in the Voigt geometry $B_{\rm V}$ ($\mathbf{B} \perp \mathbf{k}$). The coherent spin dynamics is then detected through the Faraday ellipticity of the linearly-polarized probe pulses.

A typical TRFE signal, measured at $T = 1.6$~K in $B_{\rm V} = 1$\,T, is shown in the upper part of Fig.~\ref{fig1}b. It shows oscillations with the Larmor frequency $\omega_{\rm L}$ which is a linear function of $B_{\rm V}$: $\omega_\mathrm{L}=|g|\mu_B B_{\rm V}/\hbar$. This allows us to evaluate the $g$-factor of $|g|=1.20$.  $\omega_{\rm L}$ has no offset at zero magnetic field (Fig.~S4a), which is a strong argument in favor of the presence of resident carriers in the NCs rather than of carriers bound in an exciton~\cite{Grigoryev2021}. In the latter case, an offset equal to the electron-hole exchange would be expected. General theoretical arguments and the experiments on dynamic nuclear polarization (Fig.~\ref{fig2}c, see details below) indicate that the $g$-factor has a positive sign. From its value and the strong changes of $\omega_{\rm L}$ in presence of polarized nuclei we conclude that the signal is dominated by resident holes with $g_h=+1.20$ (see Supplementary Notes S1 and S2). Note, that the stronger hole-nuclear interaction compared to the electron-nuclear one is specific for lead halide perovskite semiconductors~\cite{kirstein2021}, due to their "inverted" band structure in comparison to semiconductors like GaAs or CdTe.

The TRFE amplitude decays within a nanosecond due to spin dephasing in the ensemble of NCs showing a dispersion of Larmor frequencies. The spin dephasing time at $B_{\rm V} = 1$~T is $T_2^*=0.26$~ns, and it increases up to 0.5~ns at low magnetic fields. There it becomes limited by the hole interaction with the nuclear spin fluctuations. In stronger fields, $T_2^*$ decreases as $1/B_{\rm V}$ due to the $g_h$ dispersion  $\Delta g=0.03$ (Fig.~S4b). By using the spin inertia technique~\cite{smirnov2020} we measured the longitudinal spin relaxation time $T_1=7.6$\,$\mu$s in zero longitudinal magnetic field, which gives an estimate of the upper limit of $T_2$ in the studied NCs (Supplementary Note S4).

\subsection*{Spin mode locking in nanocrystal ensemble}

The most striking feature of the TRFE signal (blue) in Fig.~\ref{fig1}b is the presence of oscillating signal at negative time delays, before the pump. Its amplitude has a maximum at small negative delays and decays with increasing negative delay with the same dephasing time as at positive delays.
%A zoom of this signal is shown in the inset.
For such a signal to occur, optically-induced spin coherence should be present in the NCs for times exceeding the repetition period of the laser pulses $T_\text{R}=13.2$\,ns, and can therefore be obtained for spin coherence times $T_2 \geq T_\text{R}$. The same effect was observed in ensembles of singly charged (In,Ga)As/GaAs QDs with both negative and positive charging~\cite{greilich2006,VarwigPRB12,yakovlevCh6}. It is caused by synchronization of the Larmor spin precession in the ensemble of NC carriers with the periodic laser pumping, termed the spin mode locking effect~\cite{yugova2012}.

The SML effect results from the efficient accumulation of spin polarization from carrier spins which Larmor frequencies $\omega_{\rm L}$ are commensurate with the laser repetition period $T_\text{R}$. These spin precession modes fulfill the phase synchronization condition (PSC): $\omega_{\rm L}=N\omega_\text{R}=2\pi N/T_\text{R}$, with $N$ being an integer. If the single spin coherence time exceeds $T_\text{R}$, the spin polarization of these modes accumulates. Modes fulfilling the PSC are shown schematically in Fig.~\ref{fig1}c, assuming a Gaussian distribution of the Larmor frequencies centered around $N=3$ for the weight of their contribution, as shown Fig.~\ref{fig1}d by the green line with the spread of Larmor frequencies $\Delta \omega$. $\Delta \omega$ is contributed by the $g$-factor dispersion and by the nuclear spin fluctuations (Supplementary Note S4). Inside this distribution, the five essential PSC modes (red needles) are located.

The signal resulting from the sum of the five PSC modes is shown at the bottom of Fig.~\ref{fig1}c. The maxima of the PSC mode signal coincide with the pump pulse arrival times, in between the signal is reduced with a symmetry relative to the center between the pump pulses: the ensemble spin polarization dephases after a pump pulse, but then revives before the next pump. The pump pulse excites each time all modes within the Gaussian distribution, but only the PSC modes are amplified from pulse to pulse and contribute to the SML signal at negative delays. Remarkably, the spin coherence time $T_2$ of the holes in individual NCs can be determined from the ratio of the signal amplitudes at negative and positive delays. This requires a modeling approach, that we will present in the following.

\begin{figure*}[t!]
\begin{center}
\includegraphics[width=17cm]{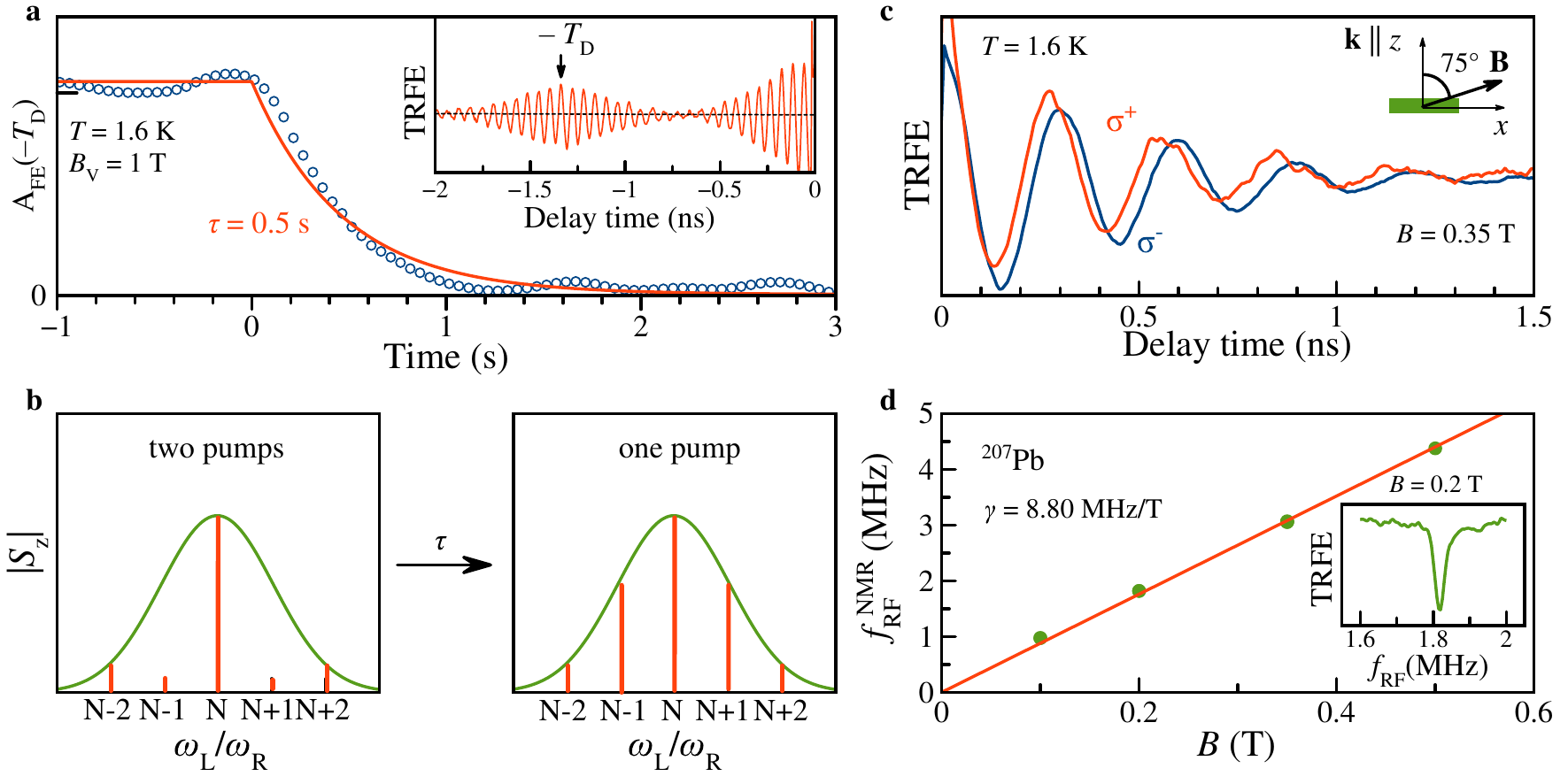}
%includegraphics[trim=0mm 0mm 0mm 0mm, clip, width=2.05\columnwidth]{Fig02.pdf}
\caption{\label{fig2} \textbf{Hole-nuclei hyperfine interaction.}
\textbf{a}, Relaxation dynamics of the burst amplitude $A_{\rm FE}(-T_\text{D})$ measured by switching from two-pump to one-pump protocol. $T =1.6$\,K and $B_{\rm V} =1$\,T. Exponential fit gives decay time of the burst amplitude $\tau=0.5$\,s, corresponding to the changeover time between the distributions, after blocking the second pump. Inset shows TRFE for negative time delays in the two-pump protocol. Arrow at delay of $-T_\text{D} = -1.32$\,ns marks the position of the amplitude relaxation measurement.
\textbf{b}, Schematic representation of the distribution $|S_z|$ of spin precession modes within the $g$-factor spread (green Gaussian) in the two-pump compared to the one-pump protocol.
\textbf{c}, TRFE signal for two circular polarizations of pump:  $\sigma^+$ (red) and $\sigma^-$ (blue). Magnetic field $B = 0.35$\,T is tilted from the light propagation direction by the angle $\alpha = 75^\circ$.
\textbf{d}, Magnetic field dependence of ODNMR frequency (dots). Line is a linear fit with slope matching the gyromagnetic ratio $\gamma = 8.8$~MHz/T of $^{207}$Pb. Inset shows typical resonance curve at $B=0.2$~T.
}
\end{center}
\end{figure*}

It was shown for (In,Ga)As QDs that the spectrum of PSC modes and the shape of TRFE signal can be tailored by using a two-pump protocol, where an additional second pump pulse is applied with a time delay $T_\text{D}$ with respect to the first pump pulse. In this case, an additional synchronization condition is introduced, as the amplified PSC modes have to satisfy two commensurability conditions: $\omega_{\rm L}=2\pi N/T_\text{R}$ and $\omega_{\rm L}=2\pi M/T_\text{D}$, with $M$ being an integer~\cite{greilich2006,yakovlevCh6}. In the experiment this leads to the emergence of signal bursts at multiple times of $T_\text{D}$. This behaviour is indeed also found in the studied perovskite NCs, as one can see from the red colored signal in Fig.~\ref{fig1}b and its zoom in Fig.~\ref{fig1}e, measured with $T_\text{D}=0.1\,T_\text{R}=1.32$\,ns. Two bursts at $-1.32$\,ns and $+2.64$\,ns are seen very prominently. A scheme of the PSC mode spectrum modification for the two-pump case is given in Fig.~\ref{fig2}b.

\subsection*{Nuclei-induced frequency focusing}

In semiconductors, the spin dynamics of electrons and holes are strongly influenced by their interaction with the nuclear spin system in which they are embedded~\cite{Glazov2018}. Polarized nuclei provide the Overhauser field $B_{\rm N}$, which shifts the carrier Larmor precession frequency to $\omega_\mathrm{L}=|g|\mu_B (B_{\rm V} + B_{\rm N})/\hbar$. Thereby, spin precession modes in Fig.~\ref{fig1}d, that do not match the PSC, can be shifted in their frequency such that they satisfy the PSC. This effect was found experimentally in (In,Ga)As QDs and identified as nuclei-induced frequency focusing (NIFF) of the carrier spin coherence~\cite{greilich2007}. The NIFF mechanism represents a positive feedback loop, that provides the required nuclear polarization in each QD, so that a large dot fraction in the ensemble is pushed into PSC modes. Due to the very slow nuclear spin relaxation at cryogenic temperatures, the NIFF configuration can be kept for hours.

In an ideal, very efficient, system as is the case for (In,Ga)As QDs with strong carrier-nuclei interaction, NIFF results in identical amplitudes of the TRFE signal at negative and positive time delays, meaning that all precession frequencies are focused on the PSC modes. This is obviously not the case for the studied perovskite NCs, see Fig.~\ref{fig1}b. Still, NIFF is present as can be evidenced by the relaxation dynamics of the bunch amplitude at $-1.32$\,ns delay when the second pump is blocked~\cite{greilich2007}. In Fig.~\ref{fig2}a we demonstrate such a measurement. We find the decay time of $\tau=0.5$\,s, which is related to the repolarization of the nuclear spin system from the two-pump to the one-pump protocol, sketched in Fig.~\ref{fig2}b. Without nuclear involvement the amplitude of the signal would decay within the carrier spin coherence time $T_2$ after switching off the second pump, as was the case for hole spins in (In,Ga)As QDs~\cite{VarwigPRB12}.

\subsection*{Dynamic nuclear polarization and ODNMR}

To further justify the involvement of nuclear spins, we apply the method of  dynamic nuclear polarization (DNP) in a titled magnetic field~\cite{kalevich_optical_1991}. The spin polarization of optically-oriented carriers is transferred via the hyperfine interaction to the nuclear spin system. The Overhauser field of the polarized nuclei $\mathbf{B}_{\rm N}$ acts back on the carriers and shifts their Larmor precession frequency, the detailed scheme can be found in Supplementary Note S2 and Ref.~\onlinecite{kirstein2021}. The shift of the Larmor frequency allows one to measure the nuclear spin polarization.

One can see in Fig.~\ref{fig2}c, that for $\sigma^+$ pump the Larmor precession becomes faster compared to the $\sigma^-$ case. The relative frequency shift is considerable, which is typical for the strong hole-nuclei spin interaction in perovskites. It allows us to confirm that the TRFE signal is related to holes and to evaluate the Overhauser field of $B_\text{N}=5.8$\,mT. The sign of the observed Larmor frequency change corresponds to a positive hole $g$-factor.

The variation of the TRFE signals by the DNP effect allows us to use it for  optical detection of nuclear magnetic resonance (ODNMR)~\cite{kirstein2021}. For that, the time delay is fixed and the TRFE amplitude is measured as a function of the radio frequency of additional radiation with $f_\text{RF}=0.1-10$~MHz. The nuclear spin system is depolarized when the energy $h f_\text{RF}$ matches to Zeeman splitting of a nuclear isotope $\mu_\text{N} g_\text{N}B$, i.e. for a NMR. Here, $\mu_\text{N}$ is the nuclear magneton and $g_\text{N}$ is the nuclear $g$-factor (Supplementary Table S1). In the experiment it is detected as a resonant decrease of the TRFE amplitude, as shown in the insert of Fig.~\ref{fig2}d at $B=0.2$\,T. From a linear fit of the resonance frequency dependence on the magnetic field $f^{\rm NMR}_\text{RF}(B)$, we evaluate the gyromagnetic ratio $\gamma = \mu_\text{N} g_\text{N}/\hbar=8.8$~MHz/T, see Fig.~\ref{fig2}d. One can conclude from the Supplementary Table S1 that among the nuclear isotopes present in the CsPb(Cl,Br)$_3$ NCs only the $^{207}$Pb isotope matches this value. The dominant role of $^{207}$Pb on the carrier spin dynamics was identified earlier in FA$_{0.9}$Cs$_{0.1}$PbI$_{2.8}$Br$_{0.2}$  crystals and the theoretical analysis establishes it as common feature for lead halide perovskites~\cite{kirstein2021}.

\subsection*{Theory of spin mode locking in perovskite NCs}

The theory of spin mode locking was previously developed for self-assembled quantum dots~\cite{YugovaPRB09} and successfully applied to explain experimental results~\cite{VarwigPRB12,yugova2012,FrasPRB2012}. However, the discovery of SML in perovskite nanocrystals provide new theoretical challenges. First, the generation of spin coherence in perovskite structures with an inverted band structure occurs on the basis of other selection rules~\cite{Glazov2018}. Circularly polarized light generates an exciton with an electron and a hole, both having spin 1/2, instead of an exciton with an electron (1/2) and a heavy hole (3/2). Second, perovskite colloidal nanocrystals in ensemble are randomly oriented. For a complete theoretical description, the anisotropy of the $g$-factor for a single NC and the possibility of generating spin coherence through positively and negatively charged trions have to be considered. Taking all these facts into account requires a significant adjustment of the theoretical model.

\begin{figure}[b!]
\begin{center}
\includegraphics[trim=0mm 0mm 0mm 0mm, clip, width=1.00\columnwidth]{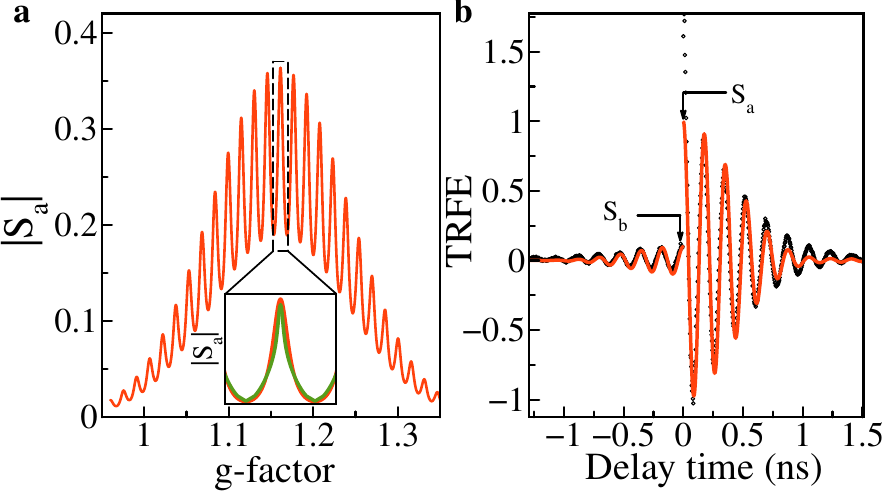}
\caption{\label{fig3} \textbf{Calculation of spin polarization and TRFE signal.}
\textbf{a}, Spectral distribution of precessing modes for $T_2 = 28$\,ns. Inset: illustration of precessing mode with NIFF for $T_2 = 13$\,ns (green) and without NIFF for $T_2 = 28$\,ns (red).
\textbf{b}, Time-resolved Faraday ellipticity (dots) measured at $B_{\rm V} = 0.35$\,T and its modeling (red line).
}
%using the parameters: $T_\text{R} = 13.2$\,ns, $\Theta = \pi$, $\Delta \omega = 2.5$\,rad/ns, $T_2 = 28$\,ns, $\omega' = 0$, $I(^{207}\text{Pb}) = 1/2$, $A_{h} = 33$\,$\mu$eV, $\alpha = 0.22$, $10^4$ nuclear spins, $\tau_c = 13$\,ns, $T_\text{1L} = 0.5$\,s, $f_\text{N} = 1$ and $\omega'\tau_p / 2\pi = 0.41$ (for NIFF). \cD{NIFF parameters are not relevant to panel (b), but only to insert to panel (a). I am not sure that we do need to give them all in main text. Probably give few important and physically trasparent, and others keep in SI?}}
\end{center}
\end{figure}

The model, which details are given in Supplementary Notes S6-S10, has two key parts: the first one considers the  spin coherence generation and its dynamics in magnetic field for a single nanocrystal with arbitrary oriented crystallographic axes. The second part accounts for inhomogeneity of the NC ensemble, namely the NCs random orientation and the $g$-factors spread.

Without loss of generality we consider resident holes, that can be created by photo-charging, as in experiment the SML of holes is observed. The resident carrier spin polarization, providing the pump-probe signal, is generated by a mechanism involving trions. The oriented spin polarization components are tilted relative to the direction of the  magnetic field, precess about it in time, and decay with the spin coherence time $T_2$. For $T_2 \geq T_R$ the spin polarization accumulates by pumping with a train of optical pulses. The SML is an ensemble effect resulting from a large number of oscillating signals with frequencies commensurate with $T_R$, therefore, the model takes into account the spread of the $g$-factors (Supplementary Note S8).

Experimental spin dynamics measured at $B_{\rm V} = 0.35$~T (black dots in Fig.~\ref{fig3}b) is modeled numerically using Supplementary Eqs.~(S28-S61). The signal decay  at negative and positive time delays is described by the $T_2^*$ that is given by the $g$ factor spread, $\Delta g$, and nuclear spin fluctuations, $\delta B_{\rm N}$.  The ratio $S_b/S_a$ of amplitudes before pulse arrival $S_b$ and after pulse arrival $S_a$ is determined by $T_2$, the optical pulse area $\Theta$ and the NIFF effect. Technically, we calculate the spin polarization distribution $S_a$ as a function of the Larmor frequency parametrized by the hole $g$ factor $\omega_{\rm L}=g_h\mu_B B_V/\hbar$ (Fig.~\ref{fig3}a). The distribution width is defined by $2\Delta g$ with $\Delta g = 0.06$. The multiple peaks correspond to the synchronized precession modes with the width of each peak determined by $1/T_2+1/\mathcal T_2^*$, where $\mathcal T_2^* \sim \hbar/(g_h\mu_B \sqrt{\delta B_{\rm N}^2})$  is related to the nuclear spin fluctuations. The modeled SML signal corresponding to the distribution is shown by red line in Fig.~\ref{fig3}b. It reproduces well all features of the experimental data.

In order to highlight the role of hole-nuclei interaction for the measured SML we perform modeling without and with account for NIFF. Without NIFF we get the hole spin coherence time $T_2 = 28$~ns for $\Theta = \pi$ as the best fit parameters. The nuclear polarization changes the spin polarization distribution by increasing $\mathcal T_2^*$ and narrowing precession modes. This results in increase of the ratio $S_b/S_a$ and to match its experimental value of 0.1 the shorter spin coherence time should be taken to keep the mode width same (insert of Fig.~\ref{fig3}a). With NIFF we get $T_2=13$~ns, which is at the lower limit of the times required for SML to occur. It is remarkable that due to the involvement of NIFF we are able to observe the SML effect for such a short-lived hole spin coherence.

\subsection*{Conclusions}

We have demonstrated the outstanding potential of perovskite NCs embedded in a glass matrix, using their spin properties as exemplary feature. We find a very long hole spin lifetime up to the microsecond range. We present an additional class of materials, in which the effects of spin mode locking and nuclei-induced frequency focusing are observed. This allows one to measure the hole spin coherence time $T_2$, disregarding the random NC orientation and variation of NC sizes and shapes. It opens the pathway for implementation of multi-pulse protocols for manipulating the spin coherence, also with involvement of strong hole-nuclei interaction provided by the inverted band structure of the lead halide perovskites. The choice of glass matrix or NC synthesis does not limit the observed effects fundamentally. We are convinced that they can be observed in many different lead halide perovskite NCs, but also in lead-free NCs.

\section*{Methods}

\textbf{Samples.}
The studied CsPb(Cl,Br)$_3$ nanocrystals embedded in fluorophosphate Ba(PO$_3$)$_2$ glass were synthesized by rapid cooling of a glass melt enriched with the components needed for the perovskite crystallization.
Samples of fluorophosphates (FP) glasses of 60Ba(PO$_3$)$_2-15$NaPO$_3-12$AlF$_3-1$Ga$_2$O$_3-4$Cs$_2$O$-8$PbF$_2$ (mol. \%) composition doped with 16 mol.\% NaCl, 3.4 mol.\% BaBr$_2$ were prepared using the melt-quench technique. The glass synthesis was performed in a closed glassy carbon crucible at temperatures of 1000-1050$^\circ$C. About 50\,g of the mixed powder was melted in the crucible during 20\,min. Then the glass melt was cast on a glassy carbon plate and pressed to form a plate with the thickness of about 2\,mm. The formation of the crystalline phase in glass can be carried out directly in the process of cooling the melt. During the pouring out of the transparent melt spontaneous precipitation of NCs with sizes smaller than the Bohr exciton radius occurs. By a subsequent heat treatment at a temperature just above the glass transition temperature ($400-429^\circ$C), it is possible to sequentially grow NCs to large sizes. The details of the method are given in Ref.~\onlinecite{kolobkova2021}.

The NC size of $7 \pm 1$~nm is evaluated from XRD data. The composition of the CsPb(Cl$_{0.5}$Br$_{0.5})_3$ NCs is estimated from the energy of the transmission edge (Fig.~\ref{fig1}a). At $T=5$\,K the exciton resonance broadened due to NC size dispersion is at 2.743~eV.  The band gap lies between that of bulk CsPbBr$_3$ and bulk CsPbCl$_3$. The optical transition is shifted by about 90~meV due to quantum confinement~\cite{krieg2021,dong2018}. For optical experiments, the sample with thickness of 500~$\mu$m was polished from both sides. The remarkable advantage of NCs in a glass matrix versus wet-chemistry synthesized NCs is the encapsulation resulting in long-term  stability. Also the possibility to have optically flat sample surfaces with diminished light scattering is important for the polarization sensitive optical techniques used in this study.

\textbf{Magneto-optical measurements.}
For low-temperature optical measurements we use a liquid helium cryostat with temperature variable from 1.6\,K up to 300\,K. At $T=1.6$\,K the sample is immersed in superfluid helium, while at $4.2-300$\,K it is held in helium gas. A superconducting vector magnet equipped with three orthogonal pairs of split coils to orient the magnetic field up to 3\,T along any arbitrary direction is used. The magnetic field parallel to $\textbf{k}$ is denoted as $B_{\rm F}$ (Faraday geometry), perpendicular to $\textbf{k}$ as $B_{\rm V}$ (Voigt geometry). If not stated otherwise, the $B_{\rm V}$ is oriented horizontally. The angle $\alpha$ is defined as the angle between $\mathbf{B}$ and the light wave vector $\mathbf{k}$, where $\alpha=0^\circ$ corresponds to $\textbf{B}_{\rm F}$.

\textbf{Transmission spectra.}
For measuring of the optical transmission a white light halogen lamp is used. The sample was thinned to 43~$\mu$m in order to avoid saturation of light absorption in the  spectral range of exciton resonances.  The spectra are detected by a 0.5\,m monochromator equipped with a silicon charge-coupled-device camera.

\textbf{Time-resolved Faraday ellipticity (TRFE).}
The coherent spin dynamics are measured using a degenerate pump-probe setup~\cite{yakovlevCh6}. A titanium-sapphire (Ti:Sa) laser generates 1.5~ps long pulses in the spectral range of $700-980$~nm ($1.265-1.771$~eV), which are frequency doubled with a beta barium borate crystal to the range of $350-490$~nm ($2.530-3.542$~eV). The laser spectral width is about 1~nm (about 2~meV), and the pulse repetition rate is 76~MHz (repetition period $T_\text{R}=13.2$~ns). The laser output is split into two beams, pump and probe, which pulses can delayed with respect to one another by a mechanical delay line. The laser photon energy is tuned to be in resonance with the exciton transition, e.g. at $2.737$~eV at $T=1.6$ and 5~K. The pump and probe beams are modulated using photo-elastic modulators (PEM). The probe beam is linearly polarized and its amplitude is modulated at 84~kHz, while the pump beam is either helicity modulated at 50~kHz between $\sigma^+/\sigma^-$ circular polarizations or amplitude modulated at 100~kHz with the circular polarization fixed at either $\sigma^+$ or $\sigma^-$. Signal was measured by lock-in amplifier at difference frequency of the pump and probe modulation. For measuring by  spin inertia  technique, the modulation frequency of the pump helicity $f_m$ was varied from 1~kHz to 5~MHz by  an electro-optical modulator (EOM). For measuring the spin dynamics induced by the pump pulses, the polarization of the transmitted probe beam was analyzed with respect to its ellipticity (Faraday ellipticity) using balanced photodiodes and a lock-in amplifier. In $B_{\rm V} \neq 0$, the Faraday ellipticity amplitude oscillates in time reflecting the Larmor precession of the charge carrier spins. The dynamics of the Faraday ellipticity signal can be described by a decaying oscillatory function:
\begin{equation}
A_{\rm FE}(t) = S \cos (\omega_{\rm L} t) \exp(-t/T^*_{2}) .
\label{eq:TRKR}
\end{equation}
Here $S$ is the amplitude at zero delay, $\omega_{\rm L}$ is the Larmor precession frequency, and $T_2^*$ the spin dephasing time. For application of a two-pump pulse scheme, after the delay line the so far not modulated linearly polarized pump beam was passed through a 50/50 nonpolarizing beam splitter. The deflected beam was sent via two mirrors over a certain distance and merged with the transmitted pump beam via a second nonpolarizing 50/50 beam splitter. By careful adjustment both pump beams were brought aligned parallel to each other, to excite the same sample spot.

\textbf{Pump-probe time-resolved differential transmission.}
Here, for excitation by the pump beam, the same scheme as in TRFE is used, but with linear polarization. The pump is amplitude modulated by a photo-elastic modulator at 100~kHz. The linearly polarized, unmodulated probe beam is separated into a reference beam, which is routed around the cryostat and the test beam, which is transmitted through the sample. The signal is recorded with respect to the intensity difference between the transmitted and the reference beams, using a balanced photodetector.

\textbf{Optically-detected nuclear magnetic resonance (ODNMR) with TRFE detection.}
In this technique nuclear magnetic resonances are measured via resonant decrease of dynamical nuclear polarization by radio frequency (RF) radiation. For that optical detection of the TRFE amplitude at fixed magnetic field is used. A small RF coil of about 5~mm diameter having 5 turns is placed close to the sample surface, similar to Ref.~\onlinecite{heisterkamp2016}. The coil is mounted flat at the sample surface and the laser beam is transmitted through the bore of the coil. The RF induced magnetic field direction is perpendicular to its surface and parallel to $\mathbf{k}$. The RF in the frequency range $f_{\rm RF}$ from 100~Hz up to 10~MHz is driven by a frequency generator, with an applied voltage of 10~V leading to an oscillating field amplitude of about 0.1~mT. The RF is terminated by internal 50$\Omega$ resistors, but not frequency matched to the circuit. In the low frequency range up to 5~MHz the current is nearly frequency independent, as the inductive resistance of the coil is small compared to the internal termination. For observation of DNP it is essential to tilt the external magnetic field away from the Voigt geometry, to have a nonzero scalar product $\mathbf{B} \cdot \mathbf{S}$ and apply a constant pump helicity.

%\textbf{Spin inertia technique.}  \DY{we need description here for PRC and spin inertia techniques to measure $T_1$. We can also make short note here that the standard approach of this techniques has been modified by us, due to the material specifics, to measure signals in both channels. The details are given in SI. } \EK{Spin inertia is shifted to SI}

%\textbf{Data availability.}
%The data on which the plots in this paper are based and other findings of our study are available from the corresponding authors upon justified request. \\

%\textbf{ORCID.} \\
%Erik Kirstein:        0000-0002-2549-2115 \\
%Dmitri R. Yakovlev:   0000-0001-7349-2745 \\
%Natalia E. Kopteva:   0000-0003-0865-0393 \\
%Evgeny A. Zhukov:     0000-0003-0695-0093 \\
%Alex Greilich:        0000-0001-7813-682X \\
%Elena V. Kolobkova:   0000-0002-0134-8434    \\
%Maria S. Kuznetsova:  0000-0003-3836-1250 \\
%Vasilii V. Belykh:    0000-0002-0032-748X \\
%Irina A. Yugova:      0000-0003-0020-3679 \\
%Mikhail M. Glazov:    0000-0003-4462-0749 \\
%Manfred Bayer:        0000-0002-0893-5949\\

\subsection*{Acknowledgements}
The authors are thankful to Al. L. Efros, M. A. Semina, and M. O. Nestoklon for fruitful discussions. We thank M. L. Skorikov for help with experiment. We acknowledge financial support by the Deutsche Forschungsgemeinschaft via the SPP2196 Priority Program (Project YA 65/26-1)  and the International Collaborative Research Centre TRR160 (Projects A1 and B2). M.M.G. acknowledges support of the Russian Foundation for Basic Research (Grant No. 19-52-12038). The low-temperature sample characterization was partially supported by the Ministry of Science and Higher Education of the Russian Federation (Contract No. 075-15-2021-598 at the P. N. Lebedev Physical Institute).  M.S.K. and I.A.Yu. acknowledge the Saint-Petersburg State University (Grant No. 91182694) and the Russian Foundation for Basic Research (RFBR Project No. 19-52-12059). Synthesis of the samples have been supported by Priority 2030 Federal Academic Leadership Program.

%\subsection*{Author contributions}
%D.R.Y., A.G. and E.K. conceived the experiment. E.K., E.A.Z. and A.G. built the experimental apparatus and performed the measurements. E.K., N.E.K., V.V.B., A.G. and D.R.Y. analyzed the data. N.E.K., I.A.Y. and M.M.G. provided the theoretical description. E.V.K. and M.S.K. fabricated and characterized the sample. All authors contributed to interpretation of the data. N.E.K., E.K., V.V.B., A.G. and D.R.Y. wrote the manuscript in close consultation with M.B.

%\subsection*{Additional information}
%Correspondence should be addressed to E.K. (email: erik.kirstein@tu-dortmund.de) and D.R.Y. (email: dmitri.yakovlev@tu-dortmund.de)

%\subsection*{Competing financial interests}
%Authors declare no competing financial interests.

%\bibliography{Perovskite_NC_Modelocking}

%\subsection*{References}

% Maximum allowed for Article 50 References.

%\bibliographystyle{naturemag}
%\normalem %to convert \emph back to italic

\end{document}

%\setcounter{equation}{0}
%\setcounter{figure}{0}
%\setcounter{table}{0}
%\setcounter{section}{0}
%\setcounter{page}{1}
%\makeatletter
%\renewcommand{\theequation}{S\arabic{equation}}
%\renewcommand{\thefigure}{S\arabic{figure}}
%\renewcommand{\bibnumfmt}[1]{[S#1]}
%\renewcommand{\citenumfont}[1]{S#1}
%\renewcommand{\thetable}{S\arabic{table}}
%\renewcommand{\thesection}{S\arabic{section}}

%\makeatletter
%\renewcommand{\fnum@figure}{Supplementary~\figurename~\thefigure}
%\makeatother	

%%%%%%%%%% Prefix a "S" to all equations, figures, tables and reset the counter %%%%%%%%%%
\setcounter{equation}{0}
\setcounter{figure}{0}
\setcounter{table}{0}
\setcounter{section}{0}
\setcounter{page}{1}
%\makeatletter
\renewcommand{\theequation}{S\arabic{equation}}
\renewcommand{\figurename}{Supplementary Figure}
\renewcommand{\thefigure}{S\arabic{figure}}
\renewcommand{\bibnumfmt}[1]{[S#1]}
\renewcommand{\citenumfont}[1]{S#1}
\renewcommand{\tablename}{Supplementary Table}
\renewcommand{\thetable}{S\arabic{table}}
\renewcommand{\thesection}{Supplementary Note S\arabic{section}}

\renewcommand{\refname}{Supplementary References}
\def\bibsection{\section*{\refname}}

%%%%%%%%%% Prefix a "S" to all equations, figures, tables and reset the counter %%%%%%%%%%

\title{Supplementary Information: \\ Mode locking of hole spin coherences in CsPb(Cl,Br)$_3$ perovskite nanocrystals \\ %in  glass matrix
}

\author{E. Kirstein, N.~E. Kopteva, D.~R. Yakovlev, E.~A. Zhukov, E.~V. Kolobkova, \\M.~S. Kuznetsova, V.~V. Belykh, I.~A. Yugova, M.~M. Glazov, M. Bayer, and A. Greilich}

%\author{E. Kirstein$^{1}$, D.~R. Yakovlev$^{1,2,3}$, N.~E. Kopteva$^{1}$, E.~A. Zhukov$^{1,2}$, E.~V. Kolobkova$^{4,5}$, M.~S. Kuznetsova$^{6}$, V.~V. Belykh$^{3}$, I.~A. Yugova$^{6}$, M.~M. Glazov$^{2}$, M.~B. Bayer$^{1,2}$, and A. Greilich$^{1}$}
%
%\affiliation{$^{1}$Experimentelle Physik 2, Department of Physics, TU Dortmund, 44227 Dortmund, Germany}
%\affiliation{$^{2}$Ioffe Institute, Russian Academy of Sciences, 194021 St. Petersburg, Russia}
%\affiliation{$^{3}$P. N. Lebedev Physical Institute of the Russian Academy of Sciences, 119991 Moscow, Russia}
%\affiliation{$^{4}$ITMO University, 199034 St. Petersburg, Russia}
%\affiliation{$^{5}$St. Petersburg State Institute of Technology, 190013 St. Petersburg, Russia}
%\affiliation{$^{6}$Spin Optics Laboratory, St. Petersburg State University, 198504 St. Petersburg, Russia}

%\date{\today}
\maketitle

\section{\label{g_factor}S1: Hole $g$-factor in nanocrystals in comparison with universal dependence on band gap for bulk perovskites}

\begin{figure}[h!]
\begin{center}
\includegraphics[width = 12cm]{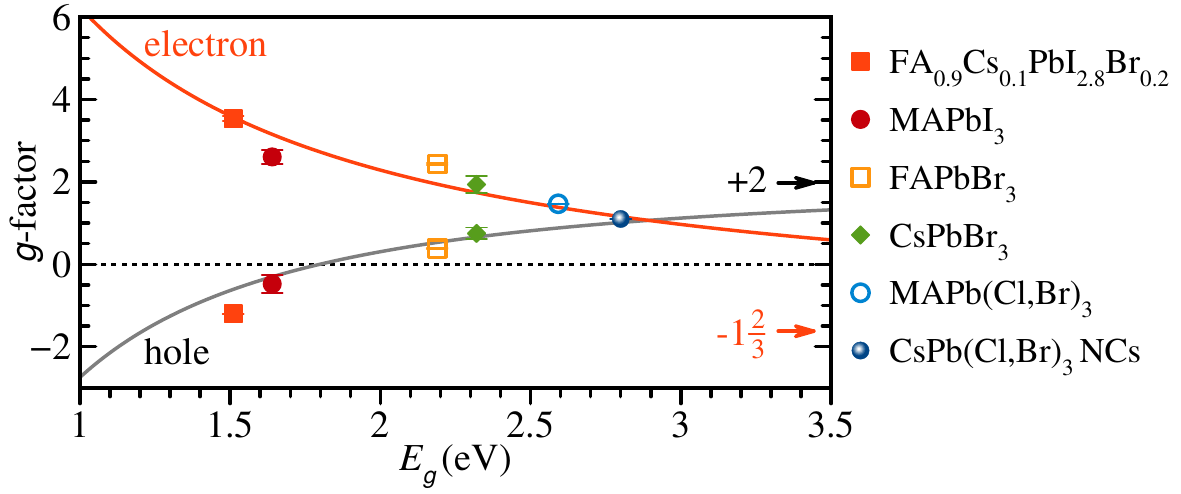}
%includegraphics[trim=0mm 0mm 0mm 0mm, clip, width=2.05\columnwidth]{Fig02.pdf}
\caption{\label{fig_g_factor} \textbf{Electron and hole $g$-factors versus band gap energy in lead halide perovskite crystals.} Experimental data for various bulk perovskites measured at $T=1.6$ and 5\,K are taken from Ref.~[\onlinecite{kirstein2021nc}] and shown by the symbols with error bars. Solid lines are model calculations, for details see Ref.~[\onlinecite{kirstein2021nc}]. The limiting values of the electron ($-5/2$) and hole ($+2$) $g$-factors for $E_g \to \infty$ are given by the arrows. Blue circle shows the hole $g$-factor in the CsPb(Cl$_{0.5}$Br$_{0.5}$)$_3$ NCs with $g_h=+1.20$, measured in the present study.}
\end{center}
\end{figure}

For convenient comparison of the $g$-factors in NCs with bulk perovskites, we use for the NCs the same approach for definition of the $g$-factor sign as in bulk, see e.g. Refs.~[\onlinecite{Ivchenko2005,Glazov2018,kirstein2021}]. In this definition the $g$-factors are positive both for electrons and holes when their ground state is the state with spin projection $S_z=-1/2$ onto the direction of the magnetic field. In this approach the exciton $g$-factor $g_X=g_e+g_h$. We used the same approach in our recent paper on electron and hole spin coherence in CsPbBr$_3$ NCs~\cite{Grigoryev2021}, while in our earlier paper~\cite{Canneson-NanoLett2017} on polarized photoluminescence in strong magnetic fields from CsPbBr$_3$ NCs another approach, common for colloidal NCs ~\cite{Efros2003}, was taken, which gives the opposite sign for the hole $g$-factor. Figure~\ref{fig_g_factor} shows the experimentally measured values of $g$-factors for electrons and holes in perovskite crystals versus the corresponding band gap energies ($E_g$). The hole $g$-factor value $g_h=+1.20$ in CsPb(Cl,Br)$_3$ NCs from the present study coincides well with the expected behavior.

%\newpage
\section{S2: Dynamic nuclear polarization in perovskites with $g_e>0$ and $g_h>0$}

\begin{figure}[h!]
\begin{center}
\includegraphics[width = 7cm]{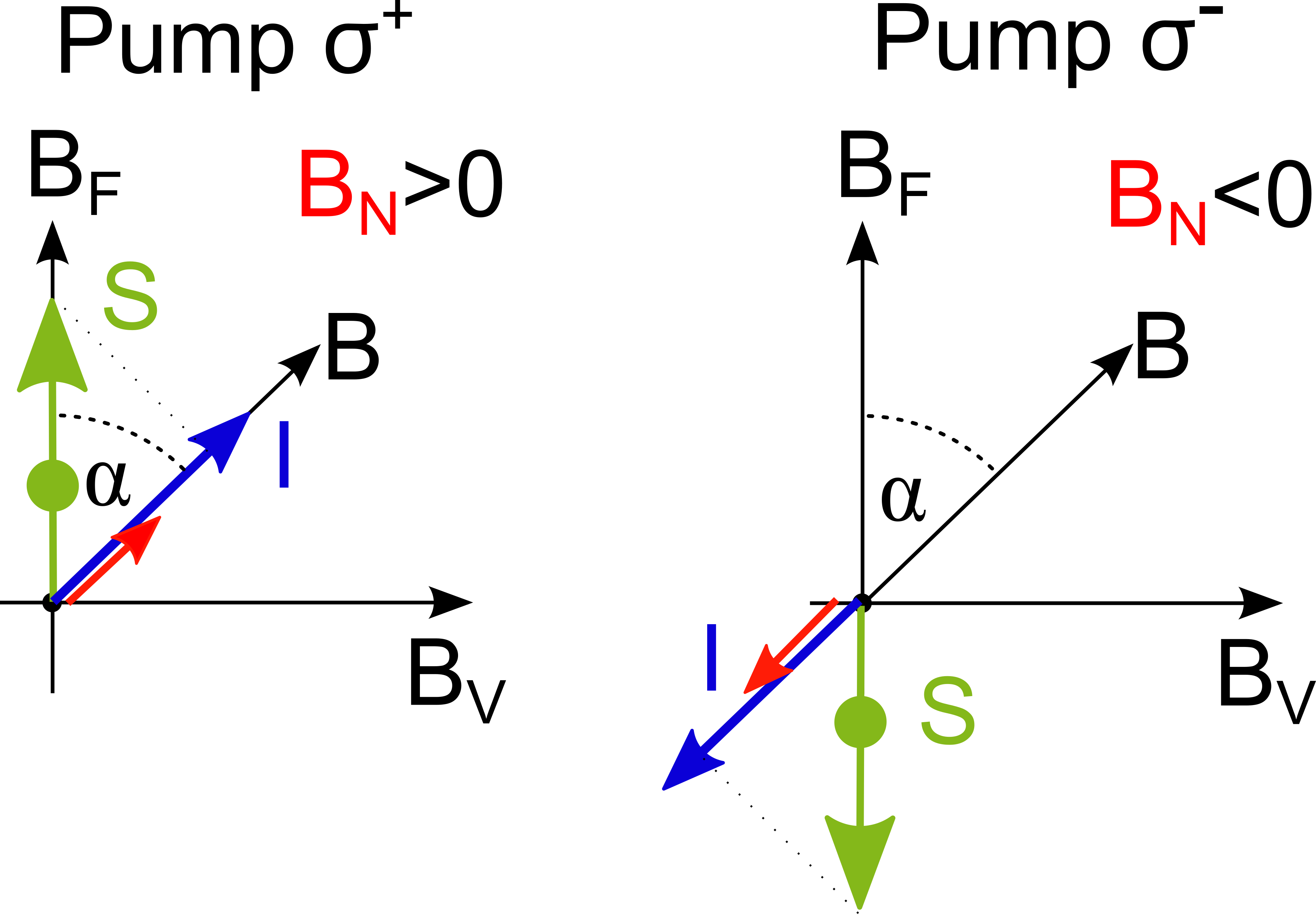}
%includegraphics[trim=0mm 0mm 0mm 0mm, clip, width=2.05\columnwidth]{Fig02.pdf}
\caption{\label{fig_DNP} \textbf{Dynamic nuclear polarization in perovskites with $g_e>0$ and $g_h>0$ in tilted magnetic field.} Scheme of interaction of charge carrier spins with nuclear spin system (identical for electrons and holes). Optically polarized carriers transfer their spin to the nuclei and induce nuclear spin polarization $\mathbf{I}$. In turn, the nuclear polarization causes an Overhauser field, $\mathbf{B}_{\rm N}$, acting back on the carriers. Orientation of the carrier spin polarization $\mathbf{S}$ (green arrow) is determined by the light helicity, $\sigma^+$ ($\sigma^-$) polarizations causes spin up (down), shown  in the left and right panels, respectively. The nuclear polarization $\mathbf{I}$ (blue arrow) builds up in the direction of $\mathbf{S}$, along the magnetic field $\mathbf{B}$ (black arrow). Here the magnetic field is inclined relative to the carrier spin polarization by an angle $\alpha$. For a $\sigma^+$ polarized pump, the nuclear spin polarization manifests as Overhauser field $\mathbf{B}_\textrm{N}$ (red arrow), which is directed along $\mathbf{I}$ and along $\mathbf{B}$. For a $\sigma^-$ polarized pump $\mathbf{B}_\textrm{N}$ and $\mathbf{B}$ are antiparallel to each other.
}
\end{center}
\end{figure}

Let us consider dynamic nuclear polarization by spin-oriented  holes. It can not be realized in the Voigt geometry, as one need to have finite projection of the hole spin polarization on the direction of external magnetic field $\mathbf{B}$ (Fig.~\ref{fig_DNP}). The circularly polarized pump pulses orient the hole spin (i.e. generate the hole spin polarization) along the direction of the k-vector of light (i.e. along the optical axis), see the green arrows in  Fig.~\ref{fig_DNP}. The hole spin polarization  $\mathbf{S}$ is transferred by flip-flop processes to the nuclear spin system. The nuclear polarization $\mathbf{I}$ is collinear with the external magnetic field $\mathbf{B}$ and its orientation is controlled  by orientation of  $\mathbf{S}$, i.e. can be controlled by the pump helicity. The oriented nuclear spins ($\mathbf{I}$) create an effective magnetic field acting back on the hole spins -- the Overhauser field ($\mathbf{B}_{\rm N}$)~\cite{kalevich_optical_1991}. If the hole spin is in a tilted external magnetic field ($\mathbf{B}$), the Overhauser field has a nonzero projection onto the direction of the external field and consequently increases or decreases the Larmor frequency of the hole spin precession. This depends on the pump circular polarization and therefore on the initial hole spin orientation, see Fig.~\ref{fig_DNP}.

Note that for CsPb(Cl,Br)$_3$ NCs, where electron and hole $g$-factors are positive ($g_e>0$ and $g_h>0$), the schemes shown in Fig.~\ref{fig_DNP} are same for the electron and hole. Change of the $g$-factor sign invert the direction of the carrier spin polarization $\mathbf{S}$ and, respectively, the direction of the nuclear polarization $\mathbf{I}$.

Which nuclear spins are affected by the carrier spins, depends on various parameters~\cite{Glazov2018}. The most important ones for the abundant non-zero nuclear spin isotopes in CsPb(Cl,Br)$_3$ are given in Tab.~\ref{tab:isotopes}.

\begin{table*}
\centering	
\caption{Major abundant non-zero nuclear spin isotopes in CsPb(Cl,Br)$_3$. The table columns give: isotope name, natural abundance $\alpha$, nuclear spin $I$, magnetic dipole moment of the isotope $\mu$ normalized to the nuclear magneton $\mu_\textrm{N}$, gyromagnetic ratio $\gamma$. Note that  $\mu=g_\textrm{N}\mu_\textrm{N}$ and $\gamma=g_\textrm{N}\mu_\textrm{N}/\hbar$, where $g_\textrm{N}$ is nuclear $g$-factor. }
\begin{tabular}{l|c|c|c|r}
			isotope & $\alpha$ & $I$ & $\mu/\mu_\mathrm{N}$ & $\gamma$ [MHz/T]  \\ \hline \hline
			$^{133}$Cs & 	100\%  &	7/2 & 2.58 & 5.623 \\
			$^{207}$Pb & 	22.1\% &	1/2 & 0.58 & 8.882 \\
			$^{35}$Cl & 	75.8\%  &	3/2 & 0.82 & 4.176 \\
			$^{37}$Cl & 	24.2\%  &	3/2 & 0.68 & 3.476 \\
			$^{79}$Br & 	50.7\%  &	3/2 & 2.10 & 10.704	\\
			$^{81}$Br & 	49.3\%  &	3/2 & 2.27 & 11.538\\

\end{tabular}
\label{tab:isotopes}
\end{table*}

%\newpage
\section{S3: Temperature dependence of hole spin mode locking}

\begin{figure}[h!]
\begin{center}
\includegraphics[width = 11cm]{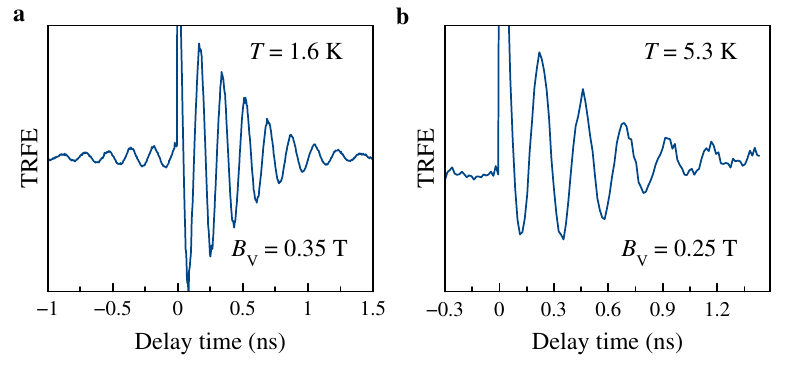}
%includegraphics[trim=0mm 0mm 0mm 0mm, clip, width=2.05\columnwidth]{Fig02.pdf}
\caption{\label{fig_temp} \textbf{Temperature dependence of hole spin mode locking.} Time-resolved Faraday ellipticity signal measured at $E_{\rm pu}=2.737$\,eV in the one-pump protocol. \textbf{a}, $B_{\rm V}=0.35$\,T and $T = 1.6$\,K. \textbf{b}, $B_{\rm V} = 0.25$\,T and $T = 5.3$\,K. Pump helicity was modulated at a frequency $f_m=10$~kHz.}
\end{center}
\end{figure}

As our previous studies show~\cite{kirstein2021}, the efficiency of the dynamic nuclear polarization strongly depends on temperature and excitation power, which should also be seen via decreased efficiency of NIFF. In the case of the studied perovskite NCs, a temperature increase from $T = 1.6$\,K to 5.3\,K results in a loss of the SML signal amplitude at negative delay times in the TRFE signal (Fig.~\ref{fig_temp}). Taking into account the strong hole localization and constant pump power, we associate this effect with reduction of the NIFF effect and/or shortening of the hole spin coherence time $T_2$. This leads to a decrease of the hole SML amplitude for $T_2 \le T_{\rm R}$, as was observed in (In,Ga)As/GaAs quantum dots~\cite{hernandezPRB08,VarwigPRB13}.

\newpage
\section{S4: Spin dephasing and spin relaxation of holes in {C\lowercase{s}P\lowercase{b}(C\lowercase{l},B\lowercase{r})}$_3$ nanocrystals}

\begin{figure}[b!]
\begin{center}
\includegraphics[width = 17cm]{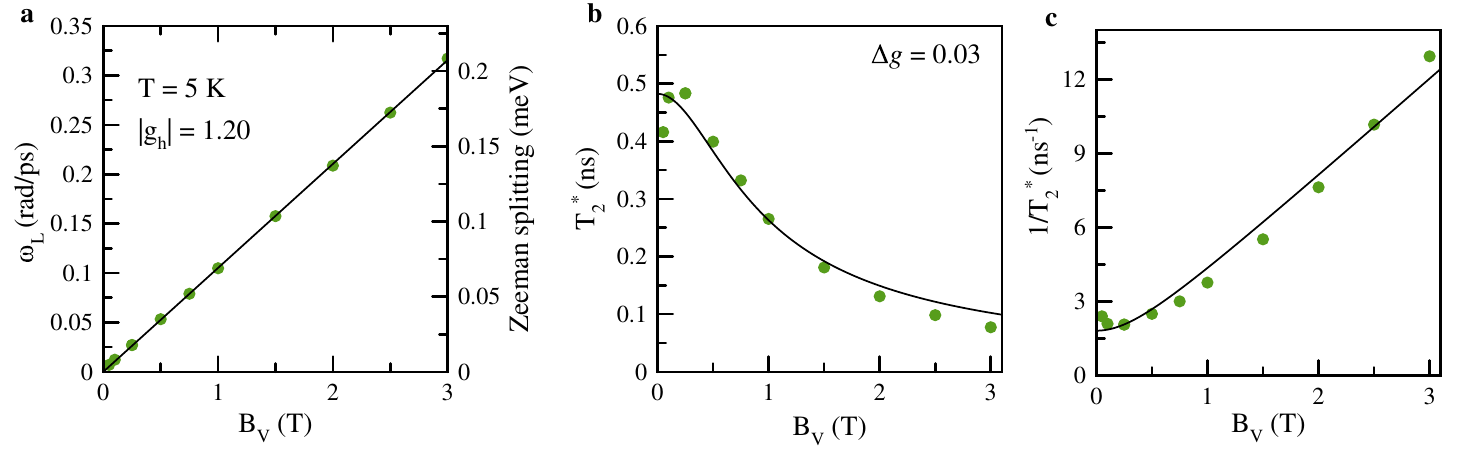}
%includegraphics[trim=0mm 0mm 0mm 0mm, clip, width=2.05\columnwidth]{Fig02.pdf}
\caption{\label{fig1} \textbf{Magnetic field dependencies.} \textbf{a}, Magnetic field dependence of the Larmor precession frequency measured at $T=5$~K for pump photon energy of 2.737~eV (green circles), $f_m=10$~kHz. Linear fit (solid line) gives $|g_h|=1.20$. \textbf{b}, Spin dephasing time $T_2^*$ and \textbf{c}, spin dephasing rate $1/T_2^*$, as functions of the magnetic field measured at $T=5$~K. Lines are fits with Eq.~(\ref{eq:intro1}), using the parameters $T^*_2(B_\text{V}=0)=0.5$\,ns and $\Delta g=0.03$.}
\end{center}
\end{figure}

Figure~\ref{fig_temp} shows the TRFE signal revealing a single-frequency damped oscillation at $T = 5.3$~K and $B_{\rm V} = 0.25$~T. The magnetic field dependence of the Larmor precession frequency $\omega_\mathrm{L}$ is shown in Fig.~\ref{fig1}\textbf{a}. The dependence of $\omega_\mathrm{L}(B_{\rm V})$ is linear without any offset at zero field, $\omega_\mathrm{L}=|g_h|\mu_B B_{\rm V}/\hbar$. It allows us to determine the hole $g$-factor value of $|g_h|=1.20$ in the CsPb(Cl$_{0.5}$Br$_{0.5}$)$_3$ NCs. The right axis in Fig.~\ref{fig1}\textbf{a} shows the value of the hole Zeeman splitting $E_Z = \hbar\omega_\mathrm{L}$. According to the theoretical dependence of the $g$-factor on the band gap, given in  Supplementary Sec.~S1, we assign a positive sign to the hole $g$-factor. The hole spin dephasing time $T_2^*$ shortens with increase of $B_{\rm V}$ due to an increased precession phase mismatch caused by the $g$-factor spread $\Delta g$ (Fig.~\ref{fig1}b). This dependence can be described by the equation:
\begin{equation}
\label{eq:intro1}
T_2^* = \frac{\hbar}{\sqrt{(\Delta g \mu_B B_{\rm V})^2 + (g_h \mu_B \Delta B)^2}},
\end{equation}
from which the spread of $g$-factors $\Delta g=0.03$ and the spin dephasing time $T_2^*(B_\text{V}=0) =\hbar / (g_h \mu_B \Delta B)= 0.5$\,ns due to the hole spin relaxation in zero external magnetic field by the nuclear spin fluctuation field $\Delta B = 20$\,mT are determined. Figure~\ref{fig1}c presents the spin dephasing rate $1/T_2^*$ as function of applied magnetic field.

\begin{figure}[t!]
\begin{center}
\includegraphics[width = 16cm]{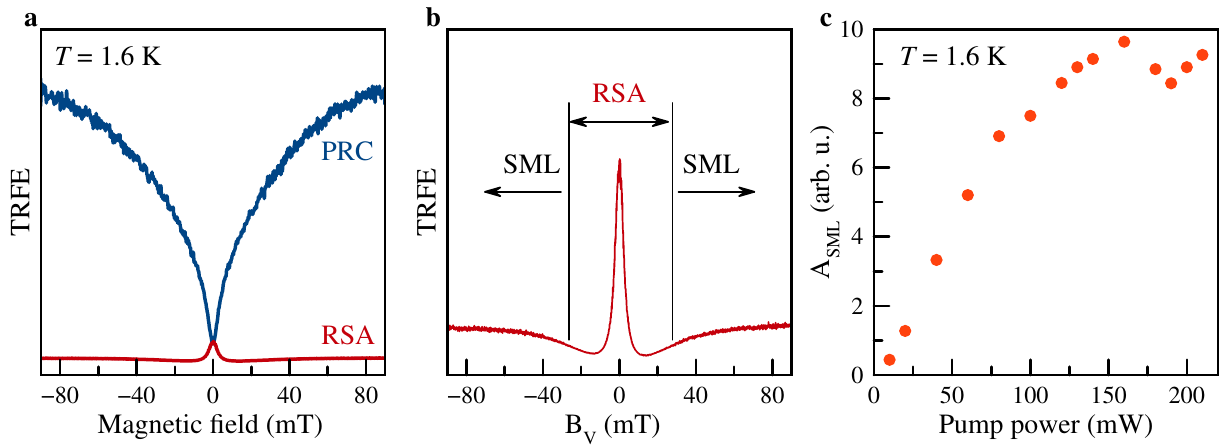}
%includegraphics[trim=0mm 0mm 0mm 0mm, clip, width=2.05\columnwidth]{Fig02.pdf}
\caption{\label{fig2n} \textbf{PRC and RSA dependencies.} \textbf{a}, PRC (blue) measured in $B_{\rm F}$, and RSA (red) measured in $B_{\rm V}$ at the negative probe delay $t=-10$~ps for $T=1.6$~K. Laser photon energy is 2.737\,eV, pump power is 20\,mW at the modulation frequency of $f_m=10$\,kHz. \textbf{b}, Zoom of the RSA signal from panel \textbf{a}. With increasing magnetic field the RSA regime is changing over to the SML regime. \textbf{c}, Spin mode locking amplitude $A_\text{SML}$ at $t<0$ as a function of the pump power measured in the one-pump protocol with $T_R = 13.2$~ns.}
\end{center}
\end{figure}

In zero magnetic field, the hole spin polarization relaxes due to interaction with random nuclear spin fluctuations. However, application of an external longitudinal magnetic field along the optical axis leads to a restoration of the spin polarization as shown in Fig.~\ref{fig2n}a by the polarization recovery curve (PRC).

The spin polarization can be decreased by a transverse magnetic field. However, due to the pulsed excitation, when the spin of a resident carrier undergoes an integer number of revolutions about the external magnetic field between subsequent pump pulse arrivals, the magnitude of spin polarization measured close to zero delay increases again - this is the case of resonant spin amplification (RSA)~\cite{yugova2012}. If the spin ensemble has a significant spread of $g$-factors, the RSA effect occurs only in a narrow range of magnetic fields around zero. Further increase of the magnetic field leads to an increasing number of possible PSC precessing modes, and the polarization magnitude becomes nonzero, reaching a constant value - the SML effect, as shown in Fig.~\ref{fig2n}a and its zoom in Fig.~\ref{fig2n}b. Such a transition of the hole spin system from RSA to SML was reported before for (In,Ga)As quantum dots with p-type doping~\cite{Varwig2012}.

The SML effect strongly depends on the pump power, as shown in  Fig.~\ref{fig2n}c. The SML amplitude increases and reaches saturation as the pump power increases. We did not observe in perovskite NCs Rabi oscillations, which have been reported for singly-charged (In,Ga)As quantum dots~\cite{Greilich2006}. Their absence is related to the pronounced inhomogeneity of the studied system.

\begin{figure}[h]
\begin{center}
\includegraphics[width = 16cm]{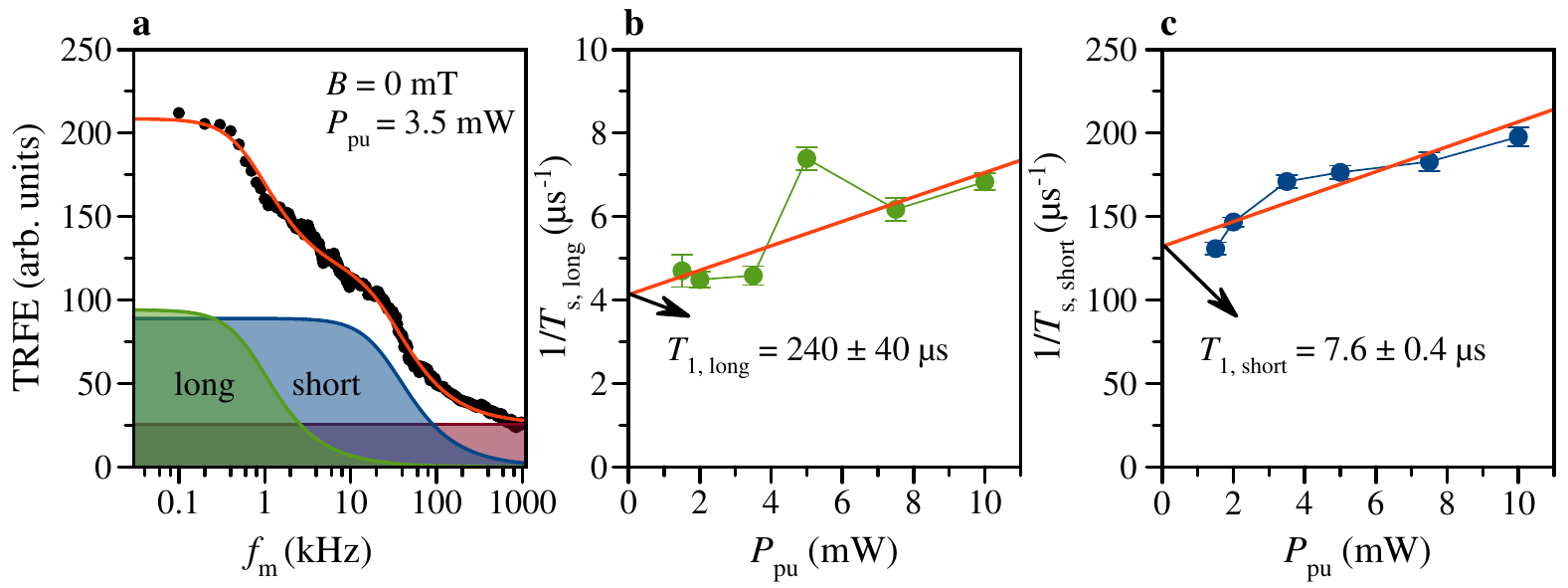}
\caption{\label{fig:T1} \textbf{Spin inertia.}
\textbf{a}, Faraday ellipticity amplitude as a function of the pump modulation frequency $f_\text{m}$ for a pump power of $P_\text{pu} = 3.5$\,mW measured at a magnetic field of $B = 0$\,mT with a pump-probe delay of $-10$\,ps (black circles). The red line is the two-component fit. The contributions of each component are shown by the green-shaded and blue-shaded areas. The violet-shaded area gives the frequency independent offset, that can be related to the scattered light.
\textbf{b--c}, The power dependence of the two corresponding inverse  effective spin lifetimes $1/T_s$. A linear extrapolation to zero power (red lines) yields $T_\text{1,long} = (240 \pm 40)\,\mu$s (\textbf{b}) and $T_\text{1,short} = (7.6 \pm 0.4)\,\mu$s (\textbf{c}). $T = 1.6$\,K. All error bars are given by the standard deviations from the fits.}
\end{center}
\end{figure}

To complete the characterization of the spin dynamics we have  measured the longitudinal spin relaxation time $T_1$, using the spin inertia technique, see Fig.~\ref{fig:T1}~\cite{heisterkamp2015,smirnov2018}. Here, we alternate the pump helicity between $\sigma^+$ and $\sigma^-$ and measure the spin polarization response with respect to the frequency of modulation, $f_\text{m}$. By increasing the frequency one can enter a regime where the modulation period is shorter than the spin relaxation time $T_1$. Thus the signal amplitude drops~\cite{heisterkamp2015}. The decay of the ellipticity upon increasing $f_\text{m}$ can be described by the dependence $S(f_\text{m}) = S_0/\sqrt{1 + (2\pi f_\text{m} T_s)^2}$, where $T_s$ is the effective spin lifetime at the corresponding pump power. The extrapolation of $T_s$ to zero power allows for extracting the intrinsic spin relaxation $T_1$ of the carriers. Figures.~\ref{fig:T1}(b) and \ref{fig:T1}(c) depict the extracted times for two components present in the spin-inertia dependence. We relate them to two subsets of carriers in the NC ensemble and focus on the regime$f_\text{m} \geq 10$\,kHz for which only the shorter living component with $T_1=7.6$\,$\mu$s contributes significantly.

%\newpage
\section{S5: Theory of spin mode locking. Introduction}
\begin{figure}[b!]
\begin{center}
\includegraphics[width = 10cm]{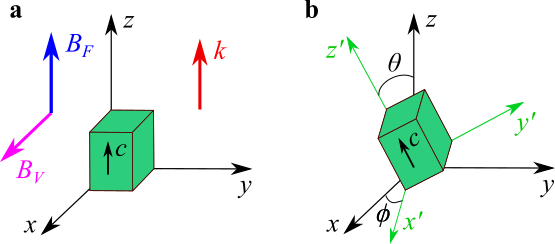}
%includegraphics[trim=0mm 0mm 0mm 0mm, clip, width=2.05\columnwidth]{FigS02.pdf}
\caption{\label{fig2} \textbf{Laboratory and NC coordinate systems used for the calculations.}
\textbf{a}, Laboratory coordinate system $(x, y, z)$ where $\mathbf{k} \parallel z$, $\mathbf{B}_{\rm F} \parallel z$ and $\mathbf{B}_{\rm V} \parallel x$ for the Faraday and Voigt geometries, respectively. The green parallelepiped symbolizes a single nanocrystal with the c-axis oriented parallel to the k-vector ($c\parallel \mathbf{k}$). \textbf{b}, Illustration of the NC coordinate frame $(x', y', z')$ tilted with respect to $(x, y, z)$. Orientation angles $\sphericalangle(z,z') \coloneqq \theta \in [0; \pi]$ and $\sphericalangle(x,x') \coloneqq \phi \in [0; 2\pi]$ are defined as shown in the scheme.}
\end{center}
\end{figure}
The theory of spin mode locking was previously developed for singly charged epitaxial quantum dots~\cite{yugova2012} and successfully applied to model experimental data in (In,Ga)As QDs~\cite{Varwig2012,Greilich2006}. However, discovering spin-locked modes in perovskite nanocrystals has provided new theoretical challenges. First, the generation of spin coherence in perovskite structures with an inverted band structure occurs on the basis of modified selection rules. Circularly polarized light generates excitons with an electron ($S=1/2$) and a hole ($S=1/2$), instead of an exciton with an electron ($S=1/2$) and a heavy hole ($S=3/2$). Second, an ensemble of perovskite colloidal nanocrystals has a random orientation of the c-axis, which leads to less efficient generation of hole spin polarization along the optical axis. The anisotropy c-axis is typical for the low temperatures when the perovskite materials expose structural transition from cubic to tetragonal or orthorhombic phase.  Third, due to the variety of perovskite materials, different nanocrystal samples may have different crystal symmetries, which is also considered in the developed model, even though the calculations are performed here only for cubic symmetry. For a complete theoretical description, the anisotropy of $g$-factors for a single NC has to be considered, as well as the possibility of generating spin coherence through positively and negatively charged trions. In order to account for all these factors  a considerable modification of the previously developed models of SML and NIFF need to be done.

The theoretical part is structured as follows:

1) Section S5 introduces the choice of the used coordinate systems.

2) In section S6 the generation of spin polarization of resident carriers (electron and hole) through intermediate trion state in a single NC is considered.

3) Section S7 considers the universal electron and hole spin polarization behavior in an external magnetic field for a single NC with an anisotropic $g$-factor.

4) Section S8 considers the inhomogeneity of the NC ensemble due to the $g$-factors dispersion and random orientation of the NC axes.

5) The modeling of the experimental results is given in Sections S9 and S10.

Let us start the presentation of the theoretical model by introducing the coordinate systems. The laboratory coordinate system is taken as $(x, y, z)$ where $\textbf{k} \parallel z$, $\mathbf{B}_{\rm F} \parallel z$ and  $\mathbf{B}_{\rm V} \parallel x$ for the Faraday and Voigt geometries, shown in Fig.~\ref{fig2}a. The single nanocrystal with $c \parallel z$ is presented by the green parallelepiped. For this NC orientation, the Voigt geometry is $\mathbf{B}_{\rm V} \parallel x$ and $\textbf{k} \parallel z$. The coordinate system $(x', y', z')$ is associated with the NC axes and $c\parallel z'$ as shown in Fig.~\ref{fig2}b. $(x', y', z')$ is tilted in respect to $(x, y, z)$ by the angles $\theta \in [0; \pi]$ and $\phi \in [0; 2\pi]$, for a NC with an arbitrary c-axis orientation. The final expressions for the spin polarization components are integrated over all possible projections of the external magnetic field and the light propagation direction on the NC axes, to simulate the random orientation of the NC axes relative to the laboratory coordinate system. One can project $(x, y, z)$ on $(x', y', z')$:
\begin{equation}\label{eq:AX01}
x' = (x\cos\theta - z\sin\theta)\cos\phi + y\sin\phi,
\end{equation}
\begin{equation}\label{eq:AX02}
y' = -(x\cos\theta - z\sin\theta)\sin\phi + y\cos\phi,
\end{equation}
\begin{equation}\label{eq:AX03}
z' = z\cos\theta + x\sin\theta.
\end{equation}

We assume that the laser light propagates along $z$ ($\textbf{k}\parallel z$), so that its electric field component $E_{z} = 0$. Therefore, the electric field components in the NC coordinate system $(x', y', z')$ are:

\begin{equation}\label{eq:AX04}
E_{x'} = E_{x}\cos\theta\cos\phi + E_{y}\sin\phi,
\end{equation}
\begin{equation}\label{eq:AX05}
E_{y'} = -E_{x}\cos\theta\sin\phi + E_{y}\cos\phi,
\end{equation}
\begin{equation}\label{eq:AX06}
E_{z'} = E_{x}\sin\theta.
\end{equation}

For the $\sigma^+$ circularly polarized light $E_{x} = i E_{y}$.

\section{S6: Optical generation of spin coherence}

\begin{figure}[t!]
\begin{center}
\includegraphics[width = 4.0cm]{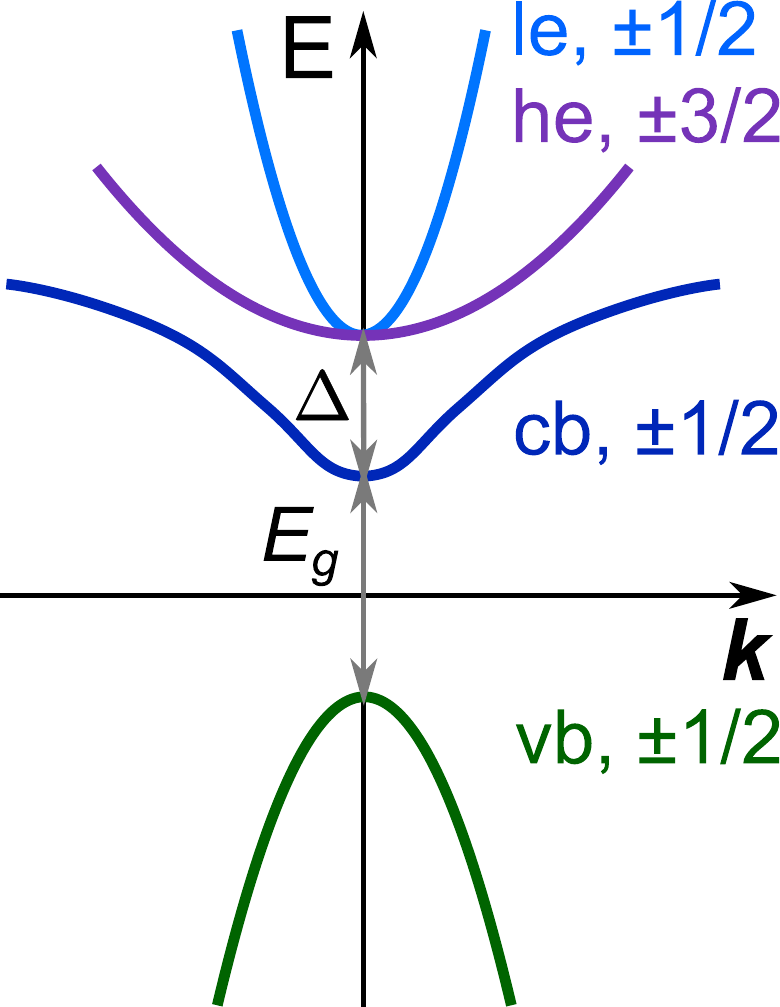}
\caption{\textbf{Schematics of perovskite band structure.} The spin $\pm 1/2$ valence band (vb) and the spin $\pm 1/2$ conduction band (cb) are separated by the optical band gap ($E_g$). Above the cb the heavy-electron (he) and light-electron (le) bands with spin $\pm3/2$ and $\pm1/2$, respectively, are located, set apart by the spin-orbit energy ($\Delta$).}
\label{figband}
\end{center}
\end{figure}

This section considers the generation of spin polarization of resident carriers (electrons and holes) in  a single NC via intermediate trion states using short optical pulses. Atomistic modeling \cite{NESTOKLON2021} shows that the $s$-orbitals of the metal (Pb) form the valence band with an admixture of the $p$-orbitals of the halogen (Br,Cl). The $p$-orbitals of the metal form the conduction band with an admixture of the halogen $s$-orbitals~\cite{kirstein2021}. Taking into account the spin-orbit interaction, the electron $\pm1/2$ spin states in the conduction band and the hole $\pm1/2$ spin states in the valence band determine the fundamental optical transitions~\cite{kirstein2021}, see also Fig.~\ref{figband}.

The matrix elements of the momentum operator $
\hat p = (p_{\perp},p_{\perp},p_{\parallel})$ taken between the conduction band Bloch functions $|c.b.,x\rangle$, $|c.b.,y\rangle$ or $|c.b.,z\rangle$  and the valence band $|v.b.\rangle$ Bloch functions at the R-point of the Brillouin zone are given by:
\begin{equation}\label{eq:OT01}
p_{\perp} = \langle c.b.,x|\hat{p}_x|v.b.\rangle = \langle c.b.,y|\hat{p}_y|v.b.\rangle, \,\,\,\,  p_{\parallel} =\langle c.b.,z|\hat{p}_z|v.b.\rangle.
\end{equation}
A representation of the wave functions is given in Ref.~\cite{kirstein2021}.

\begin{figure}[b!]
\begin{center}
\includegraphics[width = 11.0cm]{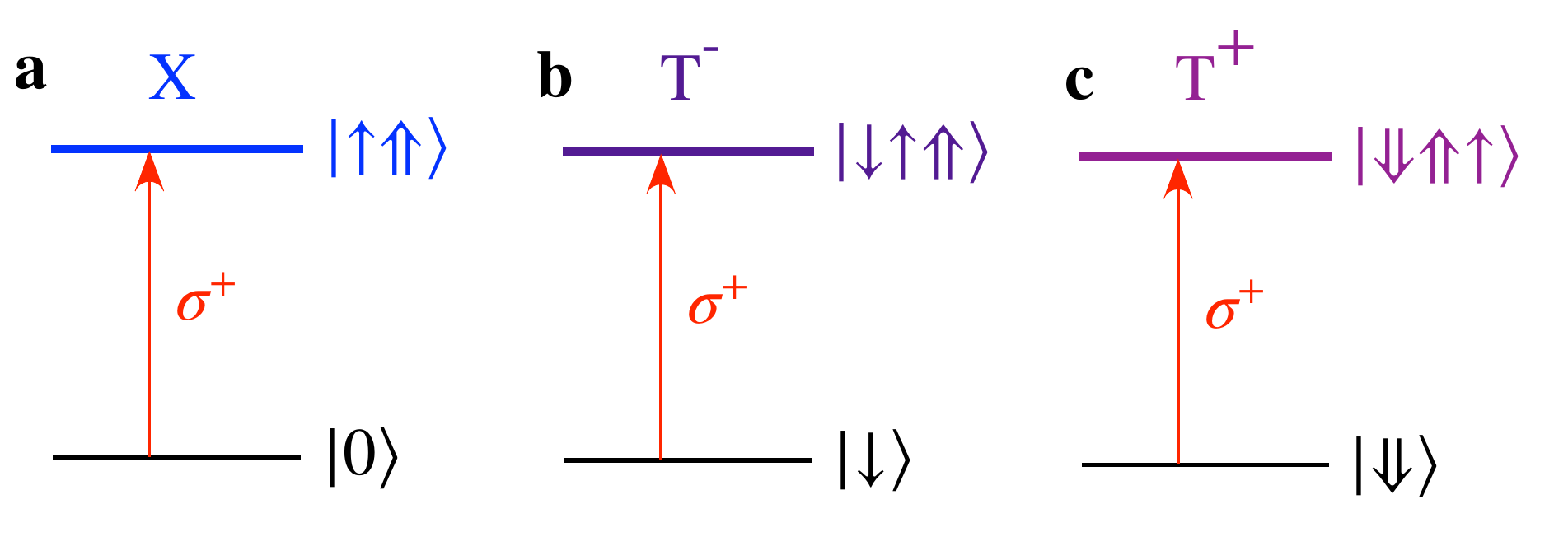}
%includegraphics[trim=0mm 0mm 0mm 0mm, clip, width=2.05\columnwidth]{Fig02.pdf}
\caption{\label{fig3N} Optical transitions for the generation by a $\sigma^+$ polarized photon of: (\textbf{a}) an exciton , (\textbf{b}) $T^-$ , and (\textbf{c}) $T^+$. $|\uparrow \rangle$ and $|\downarrow \rangle$ indicate the electron states with the wave functions $c[+1/2]$ and $c[-1/2]$, respectively. $|\Uparrow \rangle$ and $|\Downarrow \rangle$ indicate the hole states with the wave functions $v[+1/2]$ and $v[-1/2]$, respectively. The trions are photogenerated in their ground state, which is a spin singlet of the electrons or holes.}
\end{center}
\end{figure}

Pump excitation with circular polarization leads to creation of an exciton. However, as the excitons have short lifetimes ($\approx 160$\,ps), the pump-probe signal at a nanosecond time scale is provided by resident carrier spins localized in the NCs. We assume that the NCs, being initially uncharged, become charged through the optical excitation, where one of the charges becomes trapped in the NC. This photocharging in long living and, when the next optical pulse generates an additional exciton,  this leads to formation of a trion complex with the resident carrier. Figures~\ref{fig3N}b,c show level schemes of the photogeneration of negative and positive trions in a nanocrystal by $\sigma^+$ polarized excitation. If uncharged, $\sigma^+$ polarized light excites in the NC the conduction band state with wave function $c[+1/2]$ and the valence band state with wave function $v[+1/2]$, marked by $|\uparrow \rangle$ and $|\Uparrow \rangle$, respectively. If the nanocrystal contains a resident electron with wave function $c[-1/2]$, marked by $|\downarrow \rangle$, the light creates a negatively charged trion ($T^-$). In the case of a resident hole with a wave function $v[-1/2]$, marked by $|\Downarrow \rangle$, a positively charged trion ($T^+$) is generated. After the trion recombination, the spin of the resident carrier retains its polarization with a corresponding optical response in the pump-probe signal.

The matrix elements of the optical transitions created by the circularly polarized light from the electron states to the negative trion states are taken from the Ref.~\cite{Bir_Pikus}, and for the negative trion ($T^-$) can be written as:
\begin{equation}\label{eq:TT02}
M_{T^-}(c[-1/2];v[+1/2]) \propto -\frac{\cos\xi}{\sqrt{2}}(E_{x'} - iE_{y'})p_{\perp} \propto d_1 E_1,
\end{equation}
\begin{equation}\label{eq:TT01}
M_{T^-}(c[+1/2];v[-1/2]) \propto -\frac{\cos\xi}{\sqrt{2}}(E_{x'} +iE_{y'})p_{\perp} \propto d_2 E_2,
\end{equation}
\begin{equation}\label{eq:TT03}
M_{T^-}(c[-1/2];v[-1/2]) \propto -\sin\xi E_{z'}p_{\parallel} \propto d_3 E_3,
\end{equation}
\begin{equation}\label{eq:TT04}
M_{T^-}(c[+1/2];v[+1/2]) \propto \sin\xi E_{z'}p_{\parallel} \propto -d_3 E_3.
\end{equation}

One should note that for the $\sigma^+$ circularly polarized light $E_{x} = i E_{y}$ in the laboratory coordinate system then $E_{x'} \pm i E_{y'} = E_{x}\exp({\mp i\phi})(\cos\theta \mp 1)$ in the NC coordinate system.

The matrix elements for $\sigma^+$ polarized light are:
\begin{equation}\label{eq:TT041}
d_1 E_1 \propto -\frac{\cos\xi p_{\perp}}{\sqrt{2}}E_{x}\exp({i\phi})(\cos\theta + 1),
\end{equation}
\begin{equation}\label{eq:TT042}
d_2 E_2 \propto -\frac{\cos\xi p_{\perp}}{\sqrt{2}}E_{x}\exp({-i\phi})(\cos\theta - 1),
\end{equation}
\begin{equation}\label{eq:TT043}
d_3 E_3 \propto -\sin\xi p_{\parallel} E_{x}\sin\theta.
\end{equation}

In the cubic approximation $\sin\xi = 1/\sqrt{3}$, $\cos\xi = \sqrt{2/3}$ and $p_{\perp} = p_{\parallel} = p$. The matrix elements of the optical transitions to a positive trion ($T^+$) are:
\begin{equation}\label{eq:TT044}
M_{T^+}(v[-1/2];c[+1/2]) = M_{T^-}(c[-1/2];v[+1/2]) \propto d_1 E_1,
\end{equation}
\begin{equation}\label{eq:TT045}
M_{T^+}(v[+1/2];c[-1/2]) = M_{T^-}(c[+1/2];v[-1/2]) \propto d_2 E_2,
\end{equation}
\begin{equation}\label{eq:TT046}
M_{T^+}(v[+1/2];c[+1/2]) = M_{T^-}(c[-1/2];v[-1/2]) \propto d_3 E_3,
\end{equation}
\begin{equation}\label{eq:TT047}
M_{T^+}(v[-1/2];c[-1/2]) = M_{T^-}(c[+1/2];v[+1/2]) \propto -d_3 E_3.
\end{equation}

The Hamiltonian of the interaction of the nanocrystal with light in the basis ($c[1/2]$; $c[-1/2]$; $v[1/2]$; $v[-1/2]$) for $T^-$ creation is represented by:
\begin{equation}\label{eq:TT05}
\hat{H} = \frac{\hbar}{2}\begin{pmatrix}
0 & 0  & -f_3^*e^{i\omega t} & f_2^*e^{i\omega t}\\
0 & 0 & f_1^*e^{i\omega t} & f_3^*e^{i\omega t}\\
-f_3e^{-i\omega t} & f_1e^{-i\omega t} & 0 & 0\\
f_2e^{-i\omega t} & f_3e^{-i\omega t} & 0 & 0
\end{pmatrix}.
\end{equation}
Here $\hbar$ is the reduced Planck constant, $\omega$ is the optical frequency of the light, $t$ is the time, {$f_i(t) = -\frac{2e^{i\omega t}}{\hbar}\int d_i(\textbf{r})E_i(\textbf{r},t)d^3r$} is the time-dependent matrix element for description of the light interaction with a nanocrystal with index $i = (1,2,3)$. For $T^+$ in the basis ($v[1/2]$; $v[-1/2]$; $c[1/2]$; $c[-1/2]$) one needs to change the sign of $f_3$ in the Hamiltonian (\ref{eq:TT05}). One can introduce the analytical expressions for the matrix element describing the interaction with a $\sigma^+$ circularly polarized optical pulse with duration $\tau_p$ and pulse area $\Theta$:
\begin{equation}\label{eq:TT06}
f_1\tau_p \propto -\frac{\Theta}{2}e^{i\phi}(\cos\theta + 1),
\end{equation}
\begin{equation}\label{eq:TT07}
f_2\tau_p \propto \frac{\Theta}{2}e^{-i\phi}(\cos\theta - 1),
\end{equation}
\begin{equation}\label{eq:TT08}
f_3\tau_p \propto \frac{\Theta}{2}\alpha\sin\theta.
\end{equation}
Here $\alpha = p_{\parallel}\sqrt{2}\tan\xi/p_{\perp}$, in cubic approximation $\alpha = 1$. Therefore, a short pulse with rectangular shape changes the wave functions $\psi^b_1$ and $\psi^b_2$ of the electron in the conduction band $c[1/2]$ and $c[-1/2]$, respectively, to the wave functions after pulse action $\psi^a_1$ and $\psi^a_2$. Here index 'b' stands for 'before' and 'a' for 'after' pulse action:
%\newpage
%\noindent\rule{\textwidth}{1pt}
 \begin{equation}\label{eq:TT09}
\psi^a_1 = \frac{e^{i\omega'\tau_p/2}}{2}\left[\psi_1^b\left(K_p\left[1+\frac{\Delta}{W}\right] + K_m\left[1-\frac{\Delta}{W}\right]\right)+\psi_2^b\frac{2\Pi}{W}(K_p - K_m)\right],
 \end{equation}
 \begin{equation}\label{eq:TT10}
\psi^a_2 = \frac{e^{i\omega'\tau_p/2}}{2}\left[\psi_2^b\left(K_m\left[1+\frac{\Delta}
{W}\right] + K_p\left[1-\frac{\Delta}{W}\right]\right) +\psi_1^b\frac{2\Pi^*}{W}(K_p - K_m)\right].
 \end{equation}
$\omega' = \omega_p - \omega_0$ is the optical detuning between the pulse central frequency $\omega_p$ and the trion resonant frequency $\omega_0$. $\Delta$ and $\Pi$ are coefficients composed from $f_i$:
\begin{equation}\label{eq:TT13}
\Delta = \frac{1}{4}(|f_2|^2 - |f_1|^2)=-\frac{|\Theta|^2}{4}\cos\theta,
\end{equation}
\begin{equation}\label{eq:TT14}
\Pi = \frac{1}{4}(-f_1^*f_3 + f_2f_3^*)=\mp \frac{|\Theta|^2}{8} \alpha \sin \theta \exp(-i\phi).
\end{equation}
The upper sign in Eq.~(\ref{eq:TT14}) corresponds to $T^-$, and the lower one corresponds to $T^+$.
\begin{equation}\label{eq:TT15}
W = \sqrt{\Delta^2 + 4|\Pi|^2},
\end{equation}
\begin{equation}
K_{p,m} =
   \cos\left(\frac{\Omega_{p,m}\tau_p}{2}\right) - \frac{i\omega'}{\Omega_{p,m}}\sin\left(\frac{\Omega_{p,m}\tau_p}{2}\right),
\end{equation}
\begin{equation}
\Omega_{p,m} = \sqrt{(\omega')^2+\frac{1}{2}(|f_1|^2 + |f_2|^2 + 2|f_3|^2)\pm 2W}.
\end{equation}
$\Omega$ corresponds to the Rabi oscillation frequency. In the cubic approximation $\Omega\tau_p = \sqrt{(\omega'\tau_p)^2 + \Theta^2}$.

A short circularly polarized pump pulse changes the resident hole spin polarization components in a single NC according to the equations:
\begin{equation}\label{eq:SC01}
{\bf S}^a = {\bf \mathcal A} \cdot {\bf S}^b +  {\bf S}_\text{0}
\end{equation}
\begin{equation}\label{eq:SC02}
{\bf \mathcal A} = \begin{pmatrix}
\mathcal A_{11} & \mathcal A_{12} & \mathcal A_{13}\\
\mathcal A_{21} & \mathcal A_{22} &\mathcal  A_{23}\\
\mathcal A_{31} &\mathcal  A_{32} &\mathcal  A_{33}
\end{pmatrix},
\end{equation}
\begin{equation}\label{eq:SC020}
{\bf S}_\text{0} = \begin{pmatrix}
S_{x',0}\\
S_{y',0}\\
S_{z',0}
\end{pmatrix}, \hfill \hspace{1cm}
%\end{equation}
%\begin{equation}\label{eq:SC03}
{\bf S}^a = \begin{pmatrix}
S_{x'}^a \\
S_{y'}^a \\
S_{z'}^a
\end{pmatrix}, \hfill \hspace{1cm}
%\end{equation}
%\begin{equation}\label{eq:SC030}
{\bf S}^b = \begin{pmatrix}
S_{x'}^b \\
S_{y'}^b \\
S_{z'}^b
\end{pmatrix}.
\end{equation}
${\bf S}^a$ is the spin polarization after the pulse arrival, ${\bf S}^b$ is the spin polarization before the pulse arrival.
\begin{equation} \label{eq:SC04}
\mathcal A_{11} = [|K_p + K_m|^2  - \frac{(\Delta^2 - 4\text{Re}(\Pi^2))}{W^2}|K_p - K_m|^2]/4,
\end{equation}
\begin{equation} \label{eq:SC05}
\mathcal A_{12} = \frac{\Delta}{W}\text{Im}(K_pK_m^*)  - \frac{\text{Im}(\Pi^2)}{W^2}|K_p - K_m|^2,
\end{equation}
\begin{equation} \label{eq:SC06}
\mathcal A_{13} = \frac{2 \text{Im}(\Pi)}{W} \text{Im}(K_pK_m^*) + \frac{\Delta }{W^2}\text{Re}(\Pi)|K_p-K_m|^2 ,
\end{equation}
\begin{equation} \label{eq:SC061}
\mathcal A_{21} = -\frac{\Delta}{W}\text{Im}(K_pK_m^*) - \frac{\text{Im}(\Pi^2)}{W^2}|K_p - K_m|^2,
\end{equation}
\begin{equation} \label{eq:SC062}
\mathcal A_{22} = [|K_p + K_m|^2  - \frac{(\Delta^2 + 4\text{Re}(\Pi^2))}{W^2}|K_p - K_m|^2]/4,
\end{equation}
\begin{equation} \label{eq:SC063}
\mathcal A_{23} = \frac{2 Re(\Pi)}{W}\text{Im}(K_pK_m^*) - \frac{\Delta }{W^2}\text{Im}(\Pi)|K_p - K_m|^2 ,
\end{equation}
\begin{equation} \label{eq:SC07}
\mathcal  A_{31} = \frac{-2\text{Im}(\Pi)}{W}\text{Im}(K_pK_m^*) + \frac{\Delta}{W^2}\text{Re}(\Pi)|K_p - K_m|^2,
\end{equation}
\begin{equation} \label{eq:SC08}
\mathcal A_{32} = -\frac{2\text{Re}(\Pi)}{W}Im(K_pK_m^*) - \frac{\Delta}{W^2}\text{Im}(\Pi)|K_p - K_m|^2,
\end{equation}
\begin{equation} \label{eq:SC09}
\mathcal  A_{33} = [|K_p + K_m|^2 + \frac{(\Delta^2 - 4|\Pi|^2)}{W^2}|K_p - K_m|^2]/4,
\end{equation}
\begin{equation} \label{eq:SC10}
S_{x',0} = \frac{\text{Re}(\Pi)}{2W}(|K_p|^2 - |K_m|^2),
\end{equation}
\begin{equation} \label{eq:SC11}
S_{y',0} = -\frac{\text{Im}(\Pi)}{2W}(|K_p|^2 - |K_m|^2),
\end{equation}
\begin{equation} \label{eq:SC12}
S_{z',0} = \frac{\Delta}{4W}(|K_p|^2 - |K_m|^2).
\end{equation}
Here $K_0 = Ke^{i\omega'} - K^*e^{-i\omega'}$, $N = -iK_0$, $|\Gamma|^2 = (K- e^{-i\omega'})(K^*- e^{i\omega'})$ and $|\Sigma|^2 = (K + e^{-i\omega'})(K^* + e^{i\omega'})$.

Eqs.~(\ref{eq:SC01}-\ref{eq:SC12}) describe the generation of  the electron and hole spin polarization by a short optical pulse in a single NC with random c-axis orientation.

\section{S7: Spin dynamics in external magnetic field}

Here we describe the dynamics of the spin polarization in an external magnetic field. In a single NC, the hole or electron spin oriented by a circularly polarized pump pulse precesses about the external magnetic field, which can be described using a semiclassical approach \cite{Land_Lif}:
\begin{equation}\label{eq:precession}
\frac{d{\bf S}}{dt} = \left[\bm{\omega}\times {\bf S}\right].
\end{equation}
Here $\bm{\omega} = (\omega_{x'}, \omega_{y'}, \omega_{z'})$ is the Larmor precession frequency. $\omega_{i} = \mu_B g_i B_i/\hbar$, here ${i} = (x',y',z')$ and $\bm{g} = (g_{x'}, g_{y'}, g_{z'})$ is the anisotropic $g$-factor. Let us define that $g_{x'} = g_{y'} = g_{\bot}$ and $g_{z'} = g_{\parallel}$. The magnetic field components on the axis of the nanocrystal for an orientation $\mathbf{B} \parallel x$ in the laboratory coordinate system are given by:
\begin{eqnarray} \label{eq:MF0}
B_{x'} = B_{x}\cos\theta \cos\phi,\\
B_{y'} = -B_{x}\cos\theta \sin\phi,\\
B_{z'} = B_{x}\sin\theta.
\end{eqnarray}
Eq.~(\ref{eq:precession}) is written in a form that includes spin relaxation for an anisotropic $g$-factor in a single NC:
\begin{equation} \label{eq:MF1}
{\bf S}(t) = {\bf \mathcal B}(t) \cdot {\bf S}^a,
\end{equation}
\begin{equation} \label{eq:MF2}
{\bf \mathcal B}(t) = \begin{pmatrix}
\mathcal B_{11} & \mathcal B_{12}  & \mathcal B_{13}\\
\mathcal B_{21} & \mathcal B_{22} & \mathcal B_{23}\\
\mathcal B_{31} &\mathcal  B_{32} & \mathcal B_{33}
\end{pmatrix} \cdot \exp({-t/\tau_s}),
\end{equation}
\begin{equation} \label{eq:MF3}
\mathcal B_{11} = \cos(\omega_\text{L} t) + \frac{g^2_{\bot}}{\tilde{g}^2}\cos^2\theta\cos^2\phi[1 - \cos(\omega_\text{L} t)],
\end{equation}
\begin{equation} \label{eq:MF4}
\mathcal B_{12} = -\left[\frac{g^2_{\bot}}{2\tilde{g}^2}\cos^2\theta\sin2\phi\left[1-\cos(\omega_\text{L} t)\right] + \frac{g_{\parallel}}{\tilde{g}}\sin\theta\sin(\omega_\text{L} t)\right],
\end{equation}
\begin{equation} \label{eq:MF5}
\mathcal B_{13} = \frac{g_{\parallel}g_{\bot}}{2\tilde{g}^2}\cos\phi \sin2\theta\left[1-\cos(\omega_\text{L} t)\right] - \frac{g_{\bot}}{\tilde{g}}\sin\phi \cos\theta \sin(\omega_\text{L} t),
\end{equation}
\begin{equation} \label{eq:MF6}
\mathcal B_{21} = \frac{g_{\parallel}}{\tilde{g}}\sin\theta\sin(\omega_\text{L} t) - \frac{g_{\bot}^2}{2\tilde{g}^2}\cos^2\theta \sin2\phi\left[1-\cos(\omega_\text{L} t)\right],
\end{equation}
\begin{equation} \label{eq:MF7}
\mathcal  B_{22} = \cos(\omega_\text{L} t) + \frac{g_{\bot}^2}{\tilde{g}^2}\cos^2\theta \sin^2\phi\left[1 - \cos(\omega_\text{L} t)\right],
\end{equation}
\begin{equation} \label{eq:MF8}
\mathcal  B_{23} = -\frac{g_{\parallel}g_{\bot}}{2\tilde{g}^2} \sin\phi \sin2\theta\left[1-\cos(\omega_\text{L} t)\right] - \frac{g_{\bot}}{\tilde{g}}\cos\theta \cos\phi \sin(\omega_\text{L} t),
\end{equation}
\begin{equation} \label{eq:MF9}
\mathcal B_{31} = \frac{g_{\parallel}g_{\bot}}{2\tilde{g}^2}\cos\phi\sin2\theta \left[1-\cos(\omega_\text{L} t)\right] + \frac{g_{\bot}}{\tilde{g}}\cos\theta \sin\phi \sin(\omega_\text{L} t),
\end{equation}
\begin{equation} \label{eq:MF10}
\mathcal B_{32} = \frac{g_{\parallel}g_{\bot}}{2\tilde{g}^2}\sin\phi\sin2\theta\left[1-\cos(\omega_\text{L} t)\right] + \frac{g_{\bot}}{\tilde{g}}\cos\theta \cos\phi \sin(\omega_\text{L} t).
\end{equation}
\begin{equation} \label{eq:MF11}
\mathcal  B_{33} = 1 - \frac{g^2_{\bot}}{\tilde{g}^2}\cos^2\theta\left[1-\cos(\omega_\text{L} t)\right].
\end{equation}
${\bf S}(t)$ contains the time-dependent components of the electron or hole spin polarization. $\omega_\text{L} = \mu_\text{B}\tilde{g} B/\hbar$ is the Larmor precession frequency in the external magnetic field. $\tilde{g} = \sqrt{g^2_{\bot}\cos^2\theta + g^2_{\parallel}\sin^2\theta}$ is the effective $g$-factor. $\tau_s$ is the spin relaxation time. Here, the isotropic spin relaxation time is $\tau_s = T_2$. Also, we assume that the trion spin relaxation is faster than the spin relaxation of the resident carrier and since $g$-factors of the resident carrier and the unpaired carrier in the trion differ significantly, we can neglect the polarization "returning from the trion". As a result, the resident carrier gets polarized right after the trion is formed.

%Also, we assume that the trion spin relaxation is faster than the spin relaxation time of the resident carrier and, therefore, the contribution to the electron spin polarization due to trion recombination is negligible.

In the experiment, the pump pulse creates a spin polarization, which subsequently precess in the external magnetic field. If the carrier spins do not loose the polarization until the arrival of the next pulse, i.e. for $T_2 > T_\text{R}$, then spin polarization is accumulated. It is possible to model the experiment by considering an infinite number of pulses with subsequent dynamics in the magnetic field~\cite{YugovaPRB09}:
\begin{equation} \label{eq:MF11}
\textbf{S}^a = ({\bf I} - {\bf \mathcal A\mathcal B})^{-1}{\bf S}_0,
\end{equation}
with {\bf I} being the unity matrix.

\section{S8: Distribution of electron (hole) Larmor precession frequency in ensemble of NCs}

The dispersion of $g$-factors ($\Delta g$) and nuclear spin fluctuations ($\Delta \omega_\text{N}=g \mu_\text{B} \Delta B/\hbar$) broadens the spectrum of Larmor frequencies in an ensemble of nanocrystals~\cite{YugovaPRB09}. The resulting spectral distribution can be described by a Gaussian function:
\begin{equation}
\label{eq:distr01}
\rho(\omega_\text{L}) = \exp\left[ -\frac{(\omega_\text{L} - \omega_{\text{L},0})^2}{2(\Delta \omega)^2}\right].
\end{equation}
Here, $\omega_{\text{L},0}$ is the central Larmor frequency of the distribution and $\Delta \omega = \sqrt{(\Delta g \mu_\text{B}  B_{\rm V}/\hbar)^2 + (\Delta \omega_\text{N})^2}$ is the frequency dispersion. In a strong external magnetic fields exceeding the exchange field of the nuclear spin fluctuations ($B_{\rm V} \gg \Delta B$)  $\Delta \omega \approx \Delta g \mu_\text{B}  B_{\rm V}/\hbar$. Obviously, for electrons and holes $\Delta \omega$ and $\Delta B$ are different, as they have different $g$-factors and $\Delta g$, as well as different hyperfine constants describing their interaction with the nuclei~\cite{kirstein2021}.

\section{S9: Modeling of SML in perovskite NCs}\label{S:SML}

The simulation of the spin dynamics is carried out in five steps:\\
1) For a single NC the spin polarization components are calculated after the action of a pump pulse.\\
2) The precession of the spin polarization in an external magnetic field is calculated for a single NC.\\
3) The polarization components after action of an infinitely long pulse sequence are calculated.\\
4) Steps 1 -- 3 are repeated for all possible precession frequencies in the nanocrystal ensemble with certain $g$-factor dispersion.

\begin{figure}[b!]
\begin{center}
\includegraphics[trim=0mm 0mm 0mm 0mm, clip, width=0.50\columnwidth]{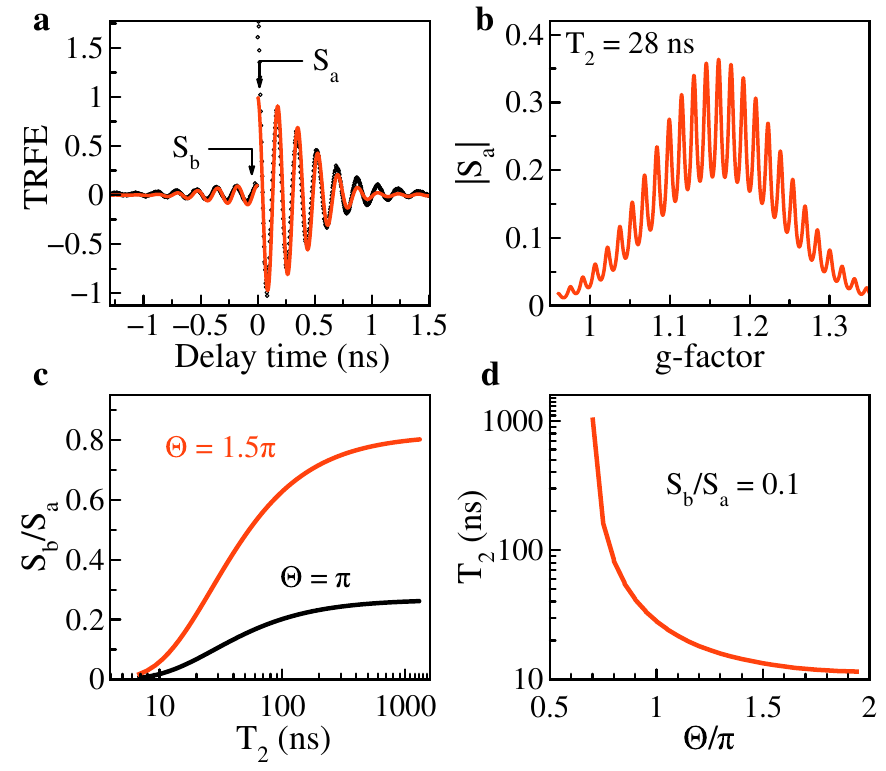}
\caption{\label{fig3} \textbf{Modeling of spin dynamics without account for NIFF.}
\textbf{a}, Time-resolved Faraday ellipticity signal (black dots) measured at $B_{\rm V} = 0.35$\,T and its modeling (red line) using parameters: $T_\text{R} = 13.2$\,ns, $\Theta = \pi$, $\Delta \omega = 2.5$\,rad/ns, $T_2 = 28$\,ns, and $\omega' = 0$.
\textbf{b}, Spectral distribution of precessing modes for $T_2 = 28$\,ns.
\textbf{c}, $S_b/S_a$ dependence on $T_2$ for $\Theta = \pi$ (black line) and $\Theta = 1.5\pi$ (red line).
\textbf{d}, Relation between the spin coherence time $T_2$ and the pulse power $\Theta$ for a fixed value of the amplitude ratio $S_b/S_a = 0.1$, as observed in the experiment.}
\end{center}
\end{figure}

Without loss of generality we consider resident holes in the NCs, that are created by photocharging. The pump-probe Faraday ellipticity signal is generated through a trion intermediate state for the resident hole spin polarization (details of spin coherence generation are given in Supplementary Note S6). The spin polarization components tilted relative to the direction of the external magnetic field precess in time and decay with the spin coherence time $T_2$. If $T_2 \geq T_R$, then the spin polarization accumulates being excited by an infinite sequence of optical pulses, as in the experiment (for details see Supplementary Note S7).

The SML is an ensemble effect being formed by adding a large number of oscillating signals with frequencies commensurate with the repetition rate of the laser pulses and $T_2 \geq T_{\rm R}$. Therefore, the model takes into account the spread of the $g$-factors (Supplementary Note S8). Numerical modeling has been performed according to Eqs.~(S26-S62). The simulated spin polarization ($S_a$) distribution after an infinite number of optical pulses at $B_{\rm V} = 0.35$~T is shown in Fig.~\ref{fig3}b. The width of the distribution is defined by $2\Delta g$ with $\Delta g = 0.06$, the multiple peaks correspond to precession modes, and the width of each peak is determined by $1/T_2$. For simplicity we assume that the spin coherence time is isotropic and the same for all precession modes. The SML signal corresponding to the distribution is shown by red line  in Fig.~\ref{fig3}a compared to the experimental dynamics (black dots). The signal decay for negative and positive time delays is defined by $\Delta g$. The ratio $S_b/S_a$ of the amplitudes before pulse arrival $S_b$ and after pulse arrival $S_a$ is determined by $T_2$, the optical pulse area $\Theta$ and also the NIFF effect which will be considered below in Supplementary Note S10. The numerical calculations show that integration over the angles $\phi$ and $\theta$ changes $\textbf{S}_a$ and $\textbf{S}_b$ in the same way. Therefore, the SML amplitude in case of cubic symmetry and the relative value $S_b/S_a$ are independent of the random orientation of the crystals. The value $S_b/S_a$ is, however, sensitive to the Rabi frequency $\Omega\tau_p = \sqrt{(\omega'\tau_p)^2 + \Theta^2}$. It also depends on the optical detuning $\omega'$ and the pump pulse duration $\tau_p$, but does not depend on $\phi$ and $\theta$. The Larmor frequency depends on the magnitude of the external magnetic field, but not on its direction.

As can be seen in Fig.~\ref{fig3}c, $S_b/S_a$ strongly depends on the spin coherence time $T_2$ due to efficiency of the spin amplification effect and on the pulse area $\Theta$ due to the Rabi frequency. The $S_b/S_a$ value increases for spin systems with longer $T_2$. The dependence of the $S_b/S_a$ value on $T_2$ reaches saturation with a value depending on the pump pulse area. Note that $\Theta = \pi$ is not optimal for the SML effect, despite that it is the optimal condition for spin coherence generation as explained in Ref.~\onlinecite{yugova2012}.

\begin{figure}[t!]
\begin{center}
\includegraphics[trim=0mm 0mm 0mm 0mm, clip, width=0.50\columnwidth]{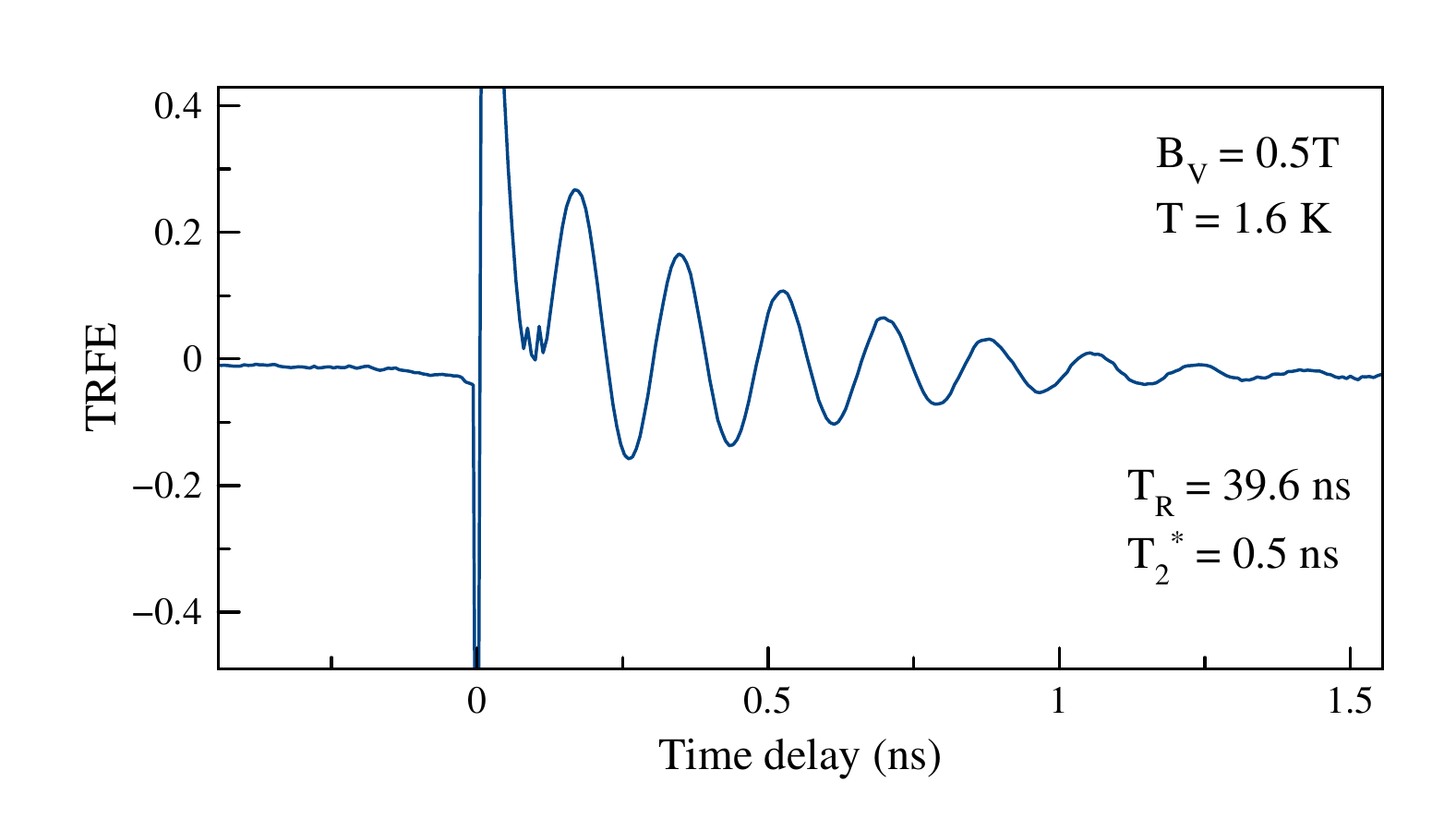}
\caption{\label{ext_PP} \textbf{Time-resolved Faraday ellipticity measured at lower repetition rate.}
Time-resolved Faraday ellipticity signal measured at $B_{\rm V} = 0.5$\,T and $T = 1.6$\,K for $T_{\rm R} = 39.6$\,ns, pump power is 3.3\,mW.}
\end{center}
\end{figure}

Figure~\ref{fig3}d shows the range of $T_2$ and $\Theta$ for which the value of $S_b/S_a = 0.1$ is taken as in the experiment. It can be obtained for $\Theta \in [0.7\pi; 2\pi]$ with $T_2$ varying from 10~ns to 1~$\mu$s. There is an experimental way to estimate the upper limit of $T_2$ time, namely to increase the period between laser pulses and reach the regime when spin coherence is fully decay before the next pump pulse arrival. In this case the SML signal amplitude at negative time delay should vanish. In Fig.~\ref{ext_PP} we show TRFE dynamics measured with a use of pulse-picker, which allows us to increase the laser repetition period up to $T_{\rm R} = 39.6$~ns. The SML signal is not detectable setting an upper limit for $T_2 < 40\,$ns. For such rather short time the SML signal generation is possible only at high pumping intensities. For $\Theta = \pi$ the hole spin coherence time of $T_2 = 28$~ns can be evaluated.

%\cD{Colleagues, for me this block was too complicated and I have changed it (see above). Please check that nothing important is missing. }
%\textit{Figure~\ref{fig3}d shows the range of $T_2$ and $\Theta$ for which the value of $S_b/S_a = 0.1$, as in the experiment. As one can see that the value $S_b/S_a = 0.1$ can be obtained for $\Theta \in [0.7\pi; 2\pi]$, with $T_2$ varying from 10\,ns to 1\,$\mu$s. However, by extending the repetition period of the laser pulses to $T_{\rm R} = 39.6$\,ns, using a pulse-picker, we demonstrate that the spin dephasing time is limited to $T_2^* = 0.5$\,ns at 3 times smaller optical excitation intensity, see Fig.~\ref{ext_PP}. As one can see at weak pumping intensity (time period between pulses is larger) the SML signal disappears, setting an upper limit $T_2 < 40\,$ns. Since we assume that the spin coherence time weakly depends on the pump intensity, the absence of signal at negative time delays shows that the spin coherence time in the system is relatively short and SML signal generation is possible only at high pump powers. This explains the choice of the parameters for the simulation: $T_2 = 28$\,ns and $\Theta = \pi$.}

%\cA{We should make a final statements of the theory part: we have SML + NIFF and $T_2 = XX$\,ns. The dispersion of different orientations does not changes anything in theory relative to QDs. Additionally, as I understand the extracted $T_2$ depends on power. In Fig.S2c we have $S_b/S_a$ ($\Theta$). Does it help us to extract the longest $T_2$? We should say somewhere that it is the longest one we can get.}

\section{S10: Evaluation of spin coherence time in the presence of NIFF}\label{S:NIFF}

Determination of the spin coherence and spin dephasing times also requires an estimate of their value in the presence of nuclei-induced frequency focusing (NIFF). After the experimental observation of the NIFF effect in (In,Ga)As quantum dots, a number of models were proposed to describe it~\cite{greilich2007,Carter2009,korenev2011,glazov2012,anders16,yaschke2017}. Among them, the model of dynamic nuclear polarization is the most physically transparent and universal one; it describes most of the observed experimental effects in various semiconductor structures~\cite{zhukov2018,evers2021a,evers2021b}. This model considers nonresonant optical pumping of the hole or electron spin ensemble and dynamic nuclear polarization arising due to the transfer of the charge carrier spin to the ensemble of nuclei via the hyperfine interaction. Such flip-flop processes are most efficient in the presence of electron or hole spin component along the external magnetic field. It was noted in Ref.~\cite{korenev2011} (see also \cite{YugovaPRB09}) that corresponding spin component $S_x$ along the magnetic field ${\bf B}_{\rm V} \parallel x$ can appear in our geometry taking into account the charge carrier spin rotation by the light pulse: In this case, there is an optical Stark field resulting in an effective magnetic field acting on the resident carrier spins that is directed along the wave vector of light. Due to the precession in this effective field, the spins gain a polarization projection ($S_x$) along the direction of the external magnetic field. The dynamics of nuclear polarization $I_{\rm N}$ along the magnetic field is described by the kinetic equation:
\begin{equation}\label{eq:DNP}
\frac{dI_{\rm N}}{dt} + \frac{1}{T_\text{1h}} [I_\text{N} - \bar{Q} \langle S_{x}(I_\text{N})\rangle] + \frac {I_\text{N}}{T_\text{d}} = 0.
\end{equation}
Here $\bar{Q} = 4I(I+1)/3$ is a factor that depends on the nuclear spin $I$ and $\bar{Q} = 1$ for perovskite NCs with $I(^{207}\text{Pb})=1/2$, $\langle S_{x}(I_\text{N})\rangle$ is the average spin polarization of the resident charge carrier along the magnetic field axis found from the solution of Eqs.~(26-61), $T_\text{1h}$ is the hyperfine coupling induced spin-flip time and $T_\text{d}$ is the nuclear spin-lattice relaxation time, which takes into account any other possible spin-leakage mechanisms. $T_\text{d} = 0.5$\,s is determined from experimental data shown in Fig.~2a in the main text. In the quasi-steady state conditions where the pulse train duration exceeds by far the nuclear relaxation times $dI_{\rm N}/dt$ can be neglected. The relative magnitude of these times can conveniently be summarized in the leakage factor $f_\text{N} = T_\text{d}/(T_\text{1h} + T_\text{d})$. The nuclear polarization produces a nonzero Overhauser field $B_{\rm N} = \alpha A_h I_\text{N}/\mu_{\rm B} g_{h}$, which acts back on the carrier spins. For our case, the hyperfine constant for the hole spins with the $^{207}\text{Pb}$ isotope having a natural abundance of $\alpha = 0.22$ is $A_h = 33$~$\mu$eV Ref.~[\onlinecite{kirstein2021}]. The nuclear polarization contributes to  the total magnetic field $B_{\rm V} + B_{\rm N}$ providing the feedback on the hole spin precession with the effective frequency $\omega_{\rm L} = \mu_{\rm B} g_{h}(B_{\rm V} + B_{\rm N})/\hbar$. Correspondingly, the hole-nuclear spin-flip rate can be estimated as
\begin{equation}\label{eq:T1N}
\frac{1}{T_\text{1h}} \propto \left(\frac{A_h}{\hbar N}\right)^2 \frac{2F\tau_{\text{c}}}{1+\omega_{\text{L}}^2\tau_{\text{c}}^2},
\end{equation}
and depends on the nuclear polarization via $\omega_{\rm L}$. Here $N$ is the number of unit cells within the hole orbit, for perovskite NCs $N = 10^4$ ~[\onlinecite{kirstein2021}], $\tau_{\text{c}}$ is the correlation time in the hole-nuclear spin system, the factor $F$ is the probability of finding the hole at the localization site. $\tau_{\text{c}} = 13$~ns is the parameter chosen to fulfill the ratio of $\omega_\text{L}\tau_{\text{c}} \gg 1$ and $F = 1$ is taken as the fitting parameter. Details of the calculation and extended equations can be found in Ref.~[\onlinecite{evers2021b}].

\begin{figure}[b!]
\begin{center}
\includegraphics[trim=0mm 0mm 0mm 0mm, clip, width=0.60\columnwidth]{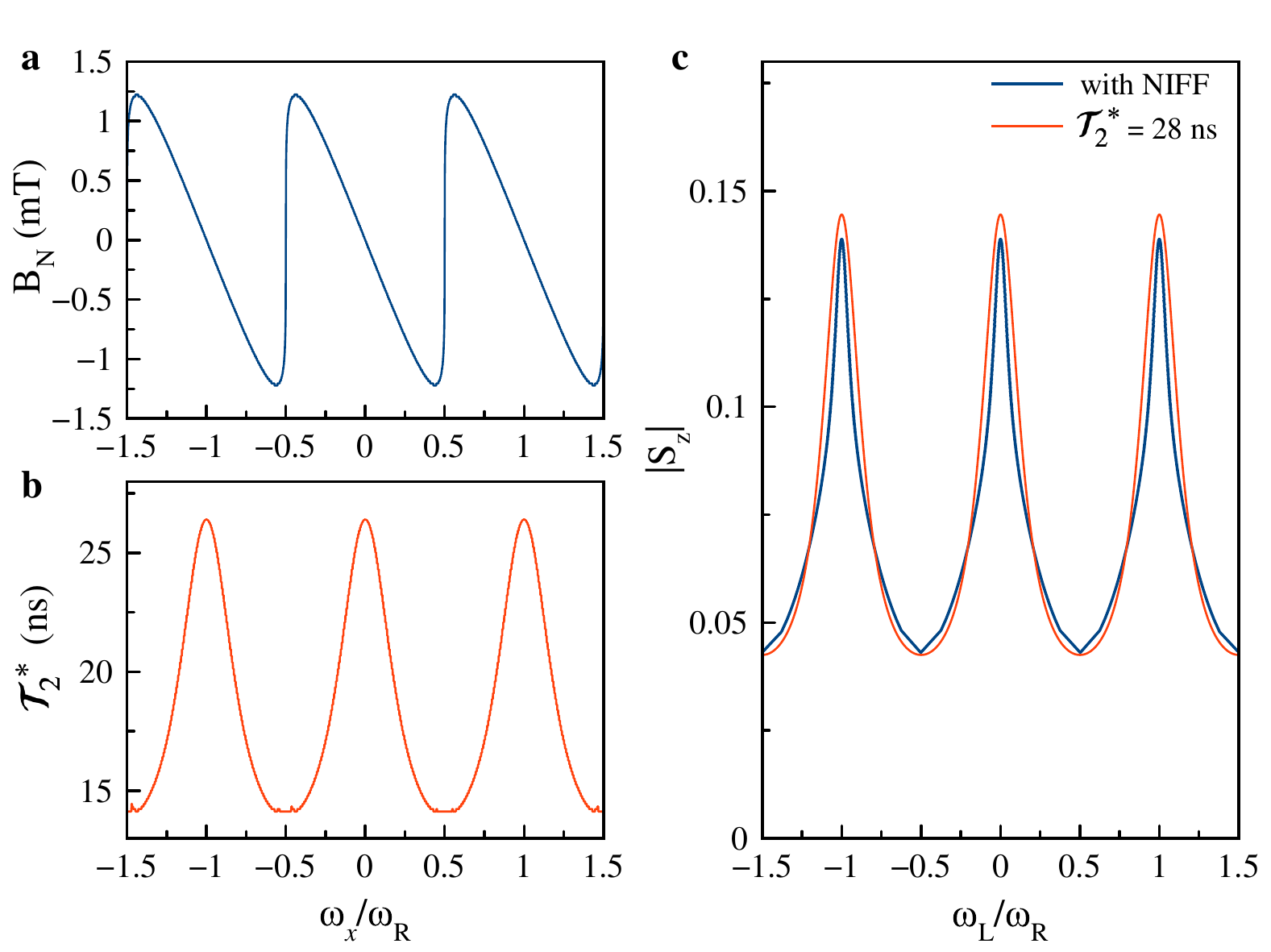}
\caption{\label{NIFF} \textbf{Nuclei-induced frequency focusing effect.}
\textbf{a}, Calculated Overhauser field distribution as a function of $\omega_x/\omega_\text{R}$ using the parameters $T_\text{R} = 13.2$\,ns, $\Theta = \pi$, $\mathcal T^{*}_{2,0} = 13$\,ns, $I(^{207}\text{Pb}) = 1/2$, $A_h = 33$~$\mu$eV, $\alpha = 0.22$, $10^4$ nuclear spins, correlation time of the hole-nuclei spin interactions $\tau_c = 13$~ns, $T_d = 0.5$~s and $\omega'\tau_p / 2\pi = 0.41$. \textbf{b}, $\mathcal T_2^*$ modulation in the presence of NIFF. \textbf{c}, Hole spin polarization distribution calculated for $\mathcal T^*_2 = 28$~ns without NIFF (red line), and in the presence of NIFF for $\mathcal T^*_{2,0} = 13$\,ns (blue line).}
\end{center}
\end{figure}

The calculated Overhauser field distribution for perovskite NCs is shown in Fig.~\ref{NIFF}a in dependence on $\omega_x/\omega_R$. The Overhauser field is $B_{\rm N} = 0$ when $\omega_x$ is commensurate with $\omega_{\rm R}$ and is maximal for hole spins precessing with frequencies commensurate with $\omega_R/2$. Therefore, all precession frequencies lying between integer values $\omega_x/\omega_{\rm R}$ are dragged towards the precession modes. The nonlinearity in the system resulting from the dependence of the hole spin precession frequency on the nuclear polarization itself results in a variation of the nuclear spin-flip rates which, in turn, affects the build-up of the nuclear spin polarization.

As a result, a non-equilibrium distribution of nuclear fields is produced. An increase of the nuclear polarization results in suppression of nuclear spin fluctuations magnitude \cite{smirnov15,zhukov2018} and extension of the resident carrier spin dephasing time:
\begin{equation}
    \mathcal T_2^* \propto \frac{\hbar}{\mu_{B}g_{{h}}\sqrt{\langle \delta B_\text{N}^2 \rangle}}.
\end{equation}
One should note that suppression of the nuclear spin fluctuations lead to distribution of the spin coherence time around the mode position, therefore we define it by homogeneous dephasing time $\mathcal T_2^*$. In the main text and in the Supplementary Note S9, distribution of the spin coherence time around the mode position is not taken into account, so that the mode width is defined by $T_2$ and the  width of the precession frequency distribution in the ensemble determines the inhomogeneous dephasing time $T_2^*$.

As suggested in Ref.~\cite{zhukov2018}, for the carrier spins satisfying the PSC
the strong feedback should lead to a reduction of the nuclear spin
fluctuations. As soon as $\omega_x/\omega_{\rm R}$ differs from $K$, the nuclear fluctuations recover due to the reduced feedback strength. Therefore, depending on the magnetic field, the spin dephasing time becomes strongly modulated due to the periodic changes of the amplitude of the nuclear fluctuations.

We calculated the spin dephasing time $\mathcal T_2^*$ dependence on magnetic field with the parameters relevant for the studied CsPb(Cl,Br)$_3$ NCs and plotted it in Fig.~\ref{NIFF}b as a function of $\omega_x/\omega_{\rm R}$.  The modulation of $\mathcal T_2^*$ around the precession mode is clearly seen. Here we have used fitting parameter $\mathcal T^*_{2,0} = 13$~ns for the maximal nuclear spin fluctuations magnitude at $\omega_x/\omega_{\rm R} = 1/2 + K$, where $K$ is an integer. The suppression of nuclear spin fluctuations leads to a prolongation of the spin dephasing time of the hole spins precessing on $\omega_x/\omega_{\rm R} = K$ as shown in Fig.~\ref{NIFF}b.

Figure~\ref{NIFF}c shows a calculated distribution near $\omega_x/\omega_{\rm R} = 0$ and $\pm 1$ to account for the spin dephasing time modulation in the dynamic nuclear polarization mechanism. The initial spin dephasing time is defined by the maximal magnitude of the nuclear spin fluctuations $\mathcal T^*_{2,0} = 13$\,ns, which becomes then extended by the reduced nuclear fluctuation at the modes. Since the calculation is numerical, we cannot give an analytical equation for the peak width, but it is comparable to the peak width for the hole polarization calculated without the NIFF effect, or for constant $\mathcal T^*_2 = T_2 = 28$\,ns.

To summarize, the effect of suppression of nuclear spin fluctuations in the dynamic nuclear polarization mechanism leads to an effective prolongation of the spin coherence time on the precession modes. This means that for a constant ratio of the mode-locking amplitudes $S_b/S_a = 0.1$ the spin coherence time extracted from the model is shorter if the NIFF effect is taken into account, $\mathcal T^*_{2,0} = T_2 =13$~ns, and longer if it is neglected, $T_2=28$~ns.

%$\mathcal T^*_{2}$ with NIFF effect determines the width of each particular mode in the distribution, and the width of the distribution as a whole is independent of the NIFF effect, so that the mode-locking signal with NIFF matches the signal modeled without involvement of nuclei.